\documentclass[%
 reprint,
 amsmath,amssymb,
 aps,
prb,
]{revtex4-2}

\usepackage{graphicx}% Include figure files
\usepackage{dcolumn}% Align table columns on decimal point
\usepackage{bm}% bold math
\usepackage[dvipsnames]{xcolor}
\usepackage[draft]{changes}% TT: switch to [final] (or remove) before submission
\usepackage{siunitx}
\usepackage{tikz}
\usepackage{multirow}
\usepackage[colorlinks=true,linkcolor=BlueViolet,citecolor=blue,urlcolor=blue]{hyperref}% add hypertext capabilities
\usepackage{glossaries}

\newacronym{tmr}{TMR}{tunnel magnetoresistance}
\newacronym{mtj}{MTJ}{magnetic tunnel junction}
\newacronym{tmdc}{TMDC}{transition metal dichalcogenide}
\newacronym{vdw}{vdW}{van der Waals}
\newacronym{sm}{SM}{Supplemental Material}
\newacronym{negf}{NEGF}{non-equilibrium Green's function}
\newacronym{ldos}{LDOS}{local density of states}

\DeclareSIUnit\Angstrom{\text{\AA}}
\glsdisablehyper

\definechangesauthor[name={Sakshi}, color=blue]{S}

\begin{document}

\title{Effects of Interfacial States and Strain on Tunnel Magnetoresistance in \\
van der Waals Magnetic Tunnel Junctions}% Force line breaks with \\

\author{Sakshi Goel}
\author{Arti Kashyap}
\email{arti@iitmandi.ac.in}
\affiliation{School of Physical Sciences, Indian Institute of Technology Mandi, Himachal Pradesh, India 175075}
\author{Keisuke Masuda}
\email{Masuda.Keisuke@nims.go.jp}
\author{Terumasa Tadano}
\email{Tadano.Terumasa@nims.go.jp}
\affiliation{National Institute for Materials Science (NIMS), 1-2-1 Sengen, Tsukuba, Ibaraki 305-0047, Japan}

\date{\today}

\begin{abstract}

All-two-dimensional magnetic tunnel junctions promise atomically sharp interfaces, yet the role of interface-induced states in their spin transport is not fully understood.
Here, we theoretically investigate spin-dependent transport in van der Waals magnetic tunnel junctions of the structure Cr$_2$C/$MY_2$/Cr$_2$C ($M$ = Mo, W; $Y$ = S, Se) with barrier thicknesses of 3, 5, 7, and 9 layers. The broad features of the $\mathbf{k}_{\parallel}$-resolved conductances, namely suppression near the $\Gamma$ point and enhancement at six off-$\Gamma$ hot spots, are consistent with the decay of evanescent states in the barrier. However, trilayer WS$_2$, MoSe$_2$, and WSe$_2$ barriers exhibit conductances of the order of $e^2/h$ at $\mathbf{k}_{\parallel}$ points within the hot spots. We attribute these near-unity transmission channels to resonant coupling between the interfacial states at the two electrode--barrier interfaces, as evidenced by their weak but finite residual weight at the barrier center.
For thicker barriers, this coupling weakens, which suppresses the residual weight, thereby reducing the tunnel magnetoresistance (TMR) ratio of the MoS$_2$ junction while enhancing those of the other junctions.
To exploit the interfacial states for spin-selective tunneling, we further examine biaxial tensile strain applied to the trilayer junctions. At 4\% strain, the TMR ratio increases from 176\% to 540\% for MoS$_2$ and from 98\% to 496\% for WS$_2$, whereas MoSe$_2$ and WSe$_2$ exhibit comparatively weaker enhancement. Our results establish interfacial-state engineering via strain and barrier thickness as effective routes for enhancing the TMR effect in all-two-dimensional magnetic tunnel junctions.
\end{abstract}

                              %display desiblue
\maketitle

%\tableofcontents

\section{Introduction}
The \gls{tmr} effect is characterized by a resistance change in \glspl{mtj} upon reversal of the relative magnetization orientation of the electrodes. In early \glspl{mtj} employing amorphous barriers, the tunneling conductance can be described in terms of the spin-resolved density of states of the electrodes, and the TMR ratio can be estimated using Julli\`{e}re's relation~\cite{Julliere1975}. A major advancement was achieved in crystalline Fe/MgO/Fe \glspl{mtj}, where tunneling is governed by symmetry-enforced spin filtering, facilitated by the matching of the majority-spin $\Delta_1$ states in Fe with the slowest-decaying evanescent state of the same symmetry in MgO~\cite{Butler2001, Mathon2001}. As a result, MgO-based MTJs demonstrated a high TMR ratio of around 200\% at room temperature (RT)~\cite{Yuasa2004, Parkin2004}. Following these initial breakthroughs, subsequent optimization of the electrode composition~\cite{DavidD2005, LeeM2006} further enhanced the performance, reaching 604\% at RT by 2008~\cite{Ikeda2008}, and more recently up to 631\% in CoFe/MgO/CoFe junctions~\cite{Scheike2023}.

Alongside these MgO-based developments, significant efforts have also been directed toward exploring alternative material systems, such as Heusler-alloy electrodes with half-metallic character~\cite{Sakuraba2006, Miura2008, Sergey2017, Aull2022}, spinel oxide barriers~\cite{Miuraspo2012, Hiroakispo2012, Sukegawa2010, Sukegawa2017, Masuda2017, Scheikespo2022}, and others~\cite{Gokaran2022, Ram2025}, in order to identify additional pathways for improving the TMR ratio.
The successful exfoliation of monolayer graphene stimulated the exploration of two-dimensional (2D) \gls{vdw} materials as tunnel barriers and further expanded the material space~\cite{Oleg2009, Piquemal2018, Piquemal-Banci2020, Kurniawan2025}.
Among 2D barriers, \glspl{tmdc} have attracted particular attention due to their tunable band gaps, controllable thickness, and compositional flexibility~\cite{Shukla2025}.
Early studies of the TMR effect in MTJs with \gls{tmdc} barriers, both experimental and theoretical, mainly employed conventional ferromagnetic (FM) electrodes such as Co~\cite{Huang2025, Shukla2025}, NiFe~\cite{WangWeiyi2015}, Fe$_3$O$_4$~\cite{Wu2015}, and Fe$_3$Si~\cite{Rotjanapittayakul2018}. In the representative Co/MoS$_2$/Co junctions, for example, theoretical calculations predicted relatively high TMR ratios, whereas the experimentally obtained values remain far below these predictions~\cite{Huang2025, Shukla2025}. This discrepancy primarily arises from the difficulty of integrating the barriers with FM electrodes while preserving the high-quality interfaces required for coherent tunneling. Even for a relatively small lattice mismatch, a perfectly coherent interface is difficult to realize~\cite{Hirohata2022}. All-2D \gls{vdw} MTJs, in which the electrodes are also layered 2D materials, can address this issue by enabling atomically sharp and clean junctions with minimal disorder~\cite{Jin2023, Lin2020}. Indeed, TMR ratios of 50\% at room temperature and 340\% at 2~K have been reported for a junction combining 2D ferromagnetic Fe$_3$GaTe$_2$ electrodes with a \gls{tmdc} barrier~\cite{Pan2024}, suggesting that similar advantages can be exploited in other all-2D junctions.

Among emerging 2D magnetic materials~\cite{Lishu2021}, MXenes have shown diverse and tunable behavior~\cite{Borge2023}. They are described by the general formula $M_{n+1}X_nT_x$ ($n = 1$--$4$), where $M$ is an early transition metal, $X$ is carbon or nitrogen, and $T_x$ denotes surface functional groups. While studies on magnetic MXenes have been largely theoretical, the recent successful fabrication of antiferromagnetic Cr$_2$N/Co bilayers via magnetron sputtering for spin--orbit torque devices highlights their practical feasibility and potential for spintronic applications~\cite{Kumar2025a, Kumar2025b}. Within this series, monolayer Cr$_2$C has been theoretically predicted to exhibit out-of-plane magnetic anisotropy and a high Curie temperature~\cite{Das2022}. The spin transport in monolayer Cr$_2$C-based MTJs has been studied with graphene and h-BN barriers~\cite{Das2022, Yu2023}, whereas junctions with \gls{tmdc} barriers remain unexplored. Notably, a strong interfacial coupling between MXene and MoS$_2$ accompanied by orbital hybridization has been reported for the Ti$_2$C/MoS$_2$ heterostructure~\cite{Gan2013}, suggesting that interface-induced states can play a crucial role in the spin transport of MXene/TMDC/MXene junctions.
It is therefore important to clarify how such interfacial states affect the spin-dependent transport in Cr$_2$C/TMDC/Cr$_2$C MTJs and to identify the conditions that maximize their TMR ratio. It should be noted that theoretically multilayer Cr$_2$C MXene exhibits an antiferromagnetic ground state~\cite{Zhangcr2cmulti2026}. In the present study, however, the ferromagnetic configuration is considered as a reference state to systematically investigate the role of interfacial states and explore their tunability for enhancing spin-dependent transport.

In this work, we investigate the spin-dependent transport in (0001)-oriented Cr$_2$C/TMDC/Cr$_2$C junctions using first-principles calculations within the \gls{negf} formalism. The \gls{tmdc} barriers considered are MoS$_2$, WS$_2$, MoSe$_2$, and WSe$_2$ with thicknesses of 3, 5, 7, and 9 layers. 
We first discuss the transport through trilayer barriers in Sec.~\ref{Subs:III(A)}, where the $\mathbf{k}_{\parallel}$-resolved conductances are interpreted in terms of the electrode band structure along the transport direction ($c$ axis) and the decay of evanescent states in the barrier. This analysis reveals a few highly conducting channels in the WS$_2$, MoSe$_2$, and WSe$_2$ junctions that cannot be explained by the barrier decay rate alone. The \gls{ldos} traces these channels to pronounced interfacial metallic states whose weight decays toward the barrier center yet remains finite there. The effects of barrier thickness and biaxial tensile strain are discussed in Secs.~\ref{Subs:III(B)} and \ref{Subs:III(C)}, respectively, where we show how these two factors modulate the induced states and thereby the TMR ratio.

\begin{figure*}
\includegraphics[width=0.8\textwidth]{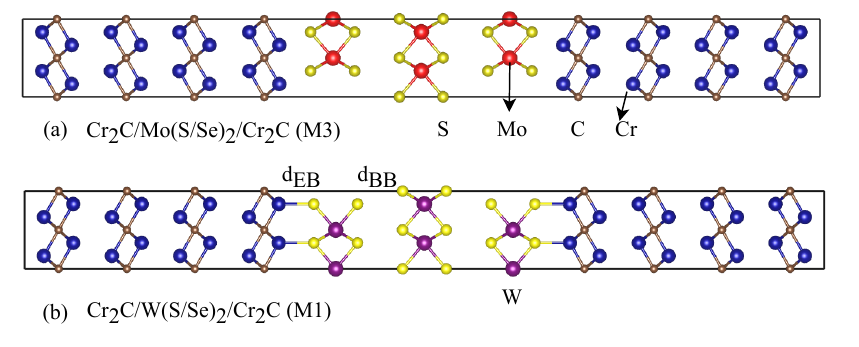}
 \caption{Scattering-region supercells of (a) Cr$_2$C/MoS$_2$/Cr$_2$C and Cr$_2$C/MoSe$_2$/Cr$_2$C MTJs (M3 stacking), and (b) Cr$_2$C/WS$_2$/Cr$_2$C and Cr$_2$C/WSe$_2$/Cr$_2$C MTJs (M1 stacking). $d_{\mathrm{EB}}$ and $d_{\mathrm{BB}}$ denote the electrode--barrier and barrier--barrier interlayer distances, respectively.}
\label{fig:mtj}
\end{figure*}

\section{Methodology}

The structural optimization and electronic-structure calculations were performed using Kohn--Sham density functional theory as implemented in QuantumATK~\cite{Smidstrup2020}. For the exchange-correlation energy, we used the Perdew--Burke--Ernzerhof~(PBE) generalized gradient approximation~(GGA) with a localized basis set, and the effect of core electrons was treated using the SG15 optimized norm-conserving Vanderbilt pseudopotentials~\cite{SG152013, Schlipf2015}.
To describe the on-site Coulomb repulsion for Cr atoms, a Hubbard $U$ parameter of 3.0~eV was used following previous studies~\cite{Yu2023, Sun2021}.
In the structural optimization step, the \gls{vdw} correction was included based on Grimme's DFT-D3 approach~\cite{Grimme2010}. A mesh cutoff of 180~Hartree was employed, and the Brillouin-zone (BZ) integration was performed with a $10 \times 10 \times 1$ Monkhorst--Pack $k$-point grid and the Fermi--Dirac broadening with an electronic temperature of \SI{300}{K}.

Monolayer Cr$_2$C MXene and 2H-\gls{tmdc} are 2D hexagonal layered materials with space groups $P\bar{3}m1$ and $P\bar{6}m2$, respectively. We first optimized the crystal structures of these monolayers using a slab model, where a sizable vacuum region of more than \SI{10}{\Angstrom} was added in the $z$ direction to avoid interactions with periodic images.
The optimized in-plane lattice constant of Cr$_2$C was 3.21~\AA, and those of the \gls{tmdc} barriers are summarized in Table~\ref{table:structure}, showing reasonably small lattice mismatches within $\sim$3\%. Next, we identified the most stable electrode/barrier interface structure by evaluating the total energies of six different stacking configurations. M3 stacking is the most stable for MoS$_2$ and MoSe$_2$, whereas M1 stacking is the most favorable for WS$_2$ and WSe$_2$. Figures~\ref{fig:mtj}(a) and \ref{fig:mtj}(b) show the M3 and M1 stackings, respectively. All six configurations are provided in the \gls{sm} (Fig.~S1, Table~S1). A supercell structure composed of four Cr$_2$C MXene layers at both ends and an $N$-layer \gls{tmdc} barrier in the center was then constructed, as shown in Fig.~\ref{fig:mtj} for the $N=3$ case. The atomic positions along the $c$ axis were optimized until the forces on each atom became less than 0.01~eV/\AA. The optimized interlayer distances between the electrode and the barrier, $d_{\mathrm{EB}}$, and those between two adjacent barrier layers, $d_{\mathrm{BB}}$, are summarized in Table~\ref{table:structure}. The calculated band gap values of the isolated monolayers are also shown and are consistent with a previous study~\cite{Deng2018}.

\begin{table}
    \caption{Structural parameters of the optimized trilayer barriers. $a$ is the in-plane lattice constant of each isolated monolayer. $d_{\mathrm{EB}}$ and $d_{\mathrm{BB}}$ are the electrode--barrier and barrier--barrier interlayer distances, respectively, computed using the heterostructure shown in Fig.~\ref{fig:mtj}. The band gap value of each isolated monolayer is also listed.}
    \label{table:structure}
\begin{ruledtabular}
\begin{tabular}{lcccc}
    Barrier & $a$ (\AA) & $d_{\mathrm{EB}}$ (\AA) & $d_{\mathrm{BB}}$ (\AA) & band gap (eV) \\
    \colrule
    MoS$_2$ & 3.184 & 2.25 & 3.23 & 1.66 \\
    WS$_2$ & 3.189 & 2.54 & 3.36 & 1.81 \\
    MoSe$_2$ & 3.319 & 2.32 & 3.16 & 1.45 \\
    WSe$_2$ & 3.318 & 2.63 & 3.36 & 1.52
\end{tabular}
\end{ruledtabular}
\end{table}

For the spin-transport calculations, the self-consistent non-equilibrium density matrix of the device region was calculated using the \gls{negf} method as implemented in QuantumATK~\cite{Smidstrup2020}. The open quantum system was constructed by combining semi-infinite Cr$_2$C electrodes with the central region. The transmission is calculated as~\cite{Brandbyge2002}
\begin{equation}
T_{\sigma}(E) = \mathrm{Tr}\!\left[\Gamma_{L}^{\sigma}(E)\,\mathcal{G}_{\sigma}^{a}(E)\,\Gamma_{R}^{\sigma}(E)\,\mathcal{G}_{\sigma}^{r}(E)\right],
\end{equation}
where $\sigma=\uparrow,\downarrow$ denotes the spin index,
$\Gamma_{L(R)}^{\sigma}$ is the coupling matrix between the left (right)
electrode and the scattering region, and
$\mathcal{G}_{\sigma}^{r(a)}$ is the retarded (advanced) Green’s function of the
scattering region that can be further written as~\cite{Smidstrup2017}
\begin{equation}
\mathcal{G}_{\sigma}^{r}(E) = \left[(E+i{\delta})S-H^{\sigma}- \Sigma_{L}^{\sigma}(E)- \Sigma_{R}^{\sigma}(E)\right]^{-1}.
\end{equation}
Here, $H$ and $S$ are the Hamiltonian and overlap matrices of the central
scattering region and
\begin{equation}
\mathcal{G}_{\sigma}^{a}(E) = \left[\mathcal{G}_{\sigma}^{r}(E)\right]^{\dagger}.
\end{equation}
$\Sigma_{L(R)}^{\sigma}$ denotes the self-energy matrix
describing the coupling of the device region to the semi-infinite left (right)
electrodes and is related to the coupling matrix via
\begin{equation}
\Gamma_{L(R)}^{\sigma}(E) = -i\left[\Sigma_{L(R)}^{\sigma}(E) - \Sigma_{L(R)}^{\sigma\,\dagger}(E)
\right].
\end{equation}
Within the Landauer formalism, the conductance of each spin channel is given by
$G_{\sigma} = (e^{2}/h)\,T_{\sigma}(E_{\mathrm{F}})$, where $E_{\mathrm{F}}$ is the Fermi energy.
Using this approach, we calculate the wave-vector-resolved conductances
$G_{\mathrm{P},\uparrow}(\mathbf{k}_{\parallel})$,
$G_{\mathrm{P},\downarrow}(\mathbf{k}_{\parallel})$,
$G_{\mathrm{AP},\uparrow}(\mathbf{k}_{\parallel})$, and
$G_{\mathrm{AP},\downarrow}(\mathbf{k}_{\parallel})$,
corresponding to the majority- and minority-spin channels in the parallel (P)
and antiparallel (AP) magnetization configurations, respectively.
The total conductances are obtained by summing over the spin channels,
\begin{equation}
G_{\mathrm{P(AP)}} = G_{\mathrm{P(AP)},\uparrow} + G_{\mathrm{P(AP)},\downarrow}.
\end{equation}
Here, $G_{\mathrm{P},\uparrow, \downarrow}$ and $G_{\mathrm{AP},\uparrow, \downarrow}$ denote the conductances averaged over the $\mathbf{k}_{\parallel}$ points in the BZ.
The \gls{tmr} ratio is defined as
\begin{equation}
\mathrm{TMR}(\%) = 100 \times\frac{G_{\mathrm{P}} - G_{\mathrm{AP}}}{G_{\mathrm{AP}}}.
\label{eq:tmr}
\end{equation}
The transmission coefficients were calculated over a
$150 \times 150 \times 1$ $\mathbf{k}_{\parallel}$ mesh, while the density-of-states
calculations for the scattering region were performed using a
% TT old text: after DFT+NEGF convergence
$61 \times 61 \times 1$ $\mathbf{k}_{\parallel}$ mesh with an infinitesimal broadening parameter of $10^{-2}$~eV.

\section{Results and Discussion}
\subsection{Trilayer barriers}\label{Subs:III(A)}
\subsubsection{\texorpdfstring{$\mathbf{k}_{\parallel}$ dependence}{k-parallel dependence}}

Table~\ref{tab:tabtri} summarizes the calculated spin-resolved conductances
$G_{\mathrm{P},\uparrow}$,
$G_{\mathrm{P},\downarrow}$,
$G_{\mathrm{AP},\uparrow}$, and
$G_{\mathrm{AP},\downarrow}$
and the TMR ratios for trilayer MoS$_2$, WS$_2$, MoSe$_2$, and WSe$_2$ barriers. 
We find that WS$_2$, MoSe$_2$, and WSe$_2$ exhibit similar
$G_{\mathrm{P},\downarrow}$ and
$G_{\mathrm{AP},\uparrow}$
values, which exceed those of MoS$_2$ by more than an order of magnitude, whereas
$G_{\mathrm{P},\uparrow}$
varies only weakly among the four barriers.

\begin{table}
\caption{Calculated spin-resolved conductances (in units of $10^{-3}\,e^2/h$) and TMR ratios (in \%) for the trilayer barriers.}
\label{tab:tabtri}
\begin{ruledtabular}
\begin{tabular}{lcccc}
$\textrm{}$&
$\textrm{MoS$_2$}$&
$\textrm{WS$_2$}$&
$\textrm{MoSe$_2$}$&
$\textrm{WSe$_2$}$\\
\colrule
$G_{\mathrm{P},\uparrow}$ & $3.68$ & $4.49$ & $11.5$ & $15.5$ \\
$G_{\mathrm{P},\downarrow}$ & $1.45$ & $55.3$ & $38.9$ & $62.0$ \\
$G_{\mathrm{AP},\uparrow}$ & $0.930$ & $15.1$ & $12.0$ & $28.5$ \\
$G_{\mathrm{AP},\downarrow}$ & $0.929$ & $15.3$ & $11.9$ & $28.7$ \\
TMR ratio & 176 & 98 & 110 & 36 \\
\end{tabular}
\end{ruledtabular}
\end{table}

\begin{figure*}
    \centering
\includegraphics[width=0.95\textwidth,clip]{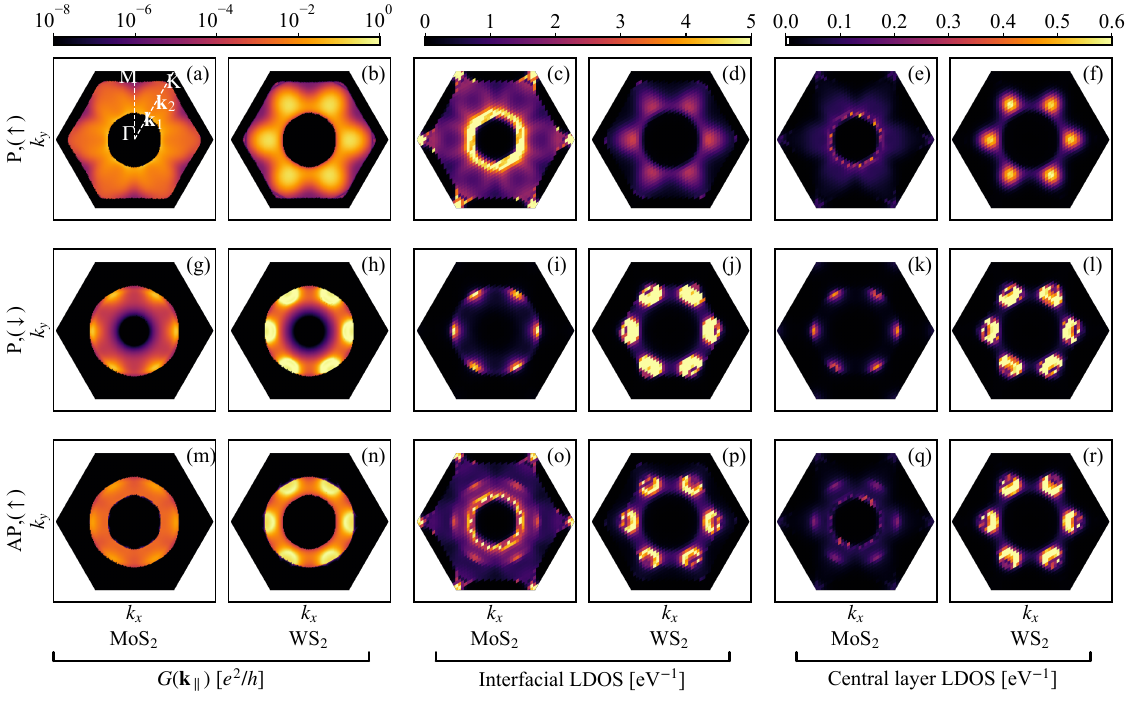}
\caption{$\mathbf{k}_{\parallel}$-resolved conductances at $E=E_\mathrm{F}$ for trilayer MoS$_2$ [(a),(g),(m)] and WS$_2$ [(b),(h),(n)]. For comparison, the corresponding LDOSs of the interfacial layers are shown in columns [(c),(i),(o)] and [(d),(j),(p)], while those of the central layer in columns [(e),(k),(q)] and [(f),(l),(r)] for MoS$_2$ and WS$_2$, respectively. The top row [(a)–(f)] corresponds to the majority-spin channel and the middle row [(g)–(l)] to the minority-spin channel in the parallel (P) magnetization configuration. The bottom row [(m)–(r)] shows the majority-spin channel in the antiparallel (AP) magnetization configuration.}
\label{fig:kp3L}
\end{figure*}

The origin of this barrier dependence can be examined in more detail by resolving the conductances over the 2D Brillouin zone. Figure~\ref{fig:kp3L} shows the $\mathbf{k}_{\parallel}=(k_x,k_y)$-resolved conductances $G_{\mathrm{P},\uparrow}(\mathbf{k}_{\parallel})$, $G_{\mathrm{P},\downarrow}(\mathbf{k}_{\parallel})$, and $G_{\mathrm{AP},\uparrow}(\mathbf{k}_{\parallel})$ at $E=E_\mathrm{F}$ for MoS$_2$ and WS$_2$. The corresponding results for MoSe$_2$ and WSe$_2$ are shown in Fig.~S2 of the SM. Because the left and right interfaces are structurally symmetric, $G_{\mathrm{AP},\uparrow}(\mathbf{k}_{\parallel})$ and $G_{\mathrm{AP},\downarrow}(\mathbf{k}_{\parallel})$ are nearly identical in both magnitude and $\mathbf{k}_{\parallel}$ dependence, and therefore only $G_{\mathrm{AP},\uparrow}(\mathbf{k}_{\parallel})$ is shown.
In the following, we denote by $\mathbf{k}_{1}$ ($\mathbf{k}_{2}$) the $\mathbf{k}_{\parallel}$ point located approximately one-third (two-thirds) of the way from $\Gamma$ to $K$.

For $G_{\mathrm{P},\uparrow}(\mathbf{k}_{\parallel})$, MoS$_2$ (MoSe$_2$) exhibits maximum-conductance regions [yellow regions in Fig.~\ref{fig:kp3L}(a)] near $\mathbf{k}_{1}$, whereas for WS$_2$ (WSe$_2$) these regions shift toward $\mathbf{k}_{2}$. Despite this shift, the magnitudes of $G_{\mathrm{P},\uparrow}(\mathbf{k}_{\parallel})$ are comparable among all TMDCs. In contrast, for both $G_{\mathrm{P},\downarrow}(\mathbf{k}_{\parallel})$ and $G_{\mathrm{AP},\uparrow}(\mathbf{k}_{\parallel})$, WS$_2$, MoSe$_2$, and WSe$_2$ exhibit broader and higher-conductance distributions than MoS$_2$ in the region around $\mathbf{k}_{2}$. Overall, all barriers exhibit very low conductance in the region centered at $\mathbf{k}_{\parallel}=\bm{0}$, specifically for $|\mathbf{k}_{\parallel}| \lesssim |\mathbf{k}_{1}|$.

Much of this $\mathbf{k}_{\parallel}$ dependence can be understood from the band structure of the Cr$_2$C electrode and the decay characteristics of the evanescent wave functions in the barrier (see \gls{sm}, Fig.~S3). Along the transport direction, minority-spin bands cross the Fermi level at $\mathbf{k}_{\parallel}=\bm{0}$, whereas majority-spin bands are absent within a disk-shaped region centered at $\mathbf{k}_{\parallel}=\bm{0}$ (radius $|\mathbf{k}_{1}|$) and appear at the Fermi level only outside this region. Despite the presence of minority-spin bands within it, the disk region contributes only weakly to $G_{\mathrm{P},\downarrow}(\mathbf{k}_{\parallel})$.

This weak contribution is explained by the decay of the evanescent states inside the barrier. Figures~S3(b)--S3(e) show the complex wave-vector decay rate ($\kappa$) at $E=E_\mathrm{F}$ across the 2D Brillouin zone for all barriers. The \gls{tmdc} barriers exhibit broadly similar $\mathbf{k}_{\parallel}$ dependencies, with moderate variations in the decay constant. In all cases, $\kappa$ is largest around $\Gamma$ and smallest in the six hot-spot regions away from it, in agreement with previous studies~\cite{Dolui2014, Shukla2025}. The larger $\kappa$ leads to stronger attenuation of the evanescent states, explaining the suppressed conductance around the $\Gamma$ point, whereas the smaller $\kappa$ values in the hot-spot regions result in weaker attenuation and hence enhanced conductance. Similar behavior has been reported for Fe and Co electrodes with multilayer \gls{tmdc} barriers irrespective of the symmetry of the electrode states along the transport direction~\cite{Dolui2014, Shukla2025}. These results indicate that the overall pattern of suppressed conductance around $\Gamma$ and enhanced conductance at the off-$\Gamma$ hot spots is governed by the $\mathbf{k}_{\parallel}$-dependent decay rate of the \gls{tmdc} barrier.

The decay rate alone, however, does not account for the differences among the barriers. Within the hot-spot regions, where all barriers exhibit only small variations in $\kappa$, MoS$_2$ shows transmission lower than that of the other TMDCs by approximately one order of magnitude, specifically for $G_{\mathrm{P},\downarrow}(\mathbf{k}_{\parallel})$ and $G_{\mathrm{AP},\uparrow}(\mathbf{k}_{\parallel})$. Moreover, as shown in Figs.~\ref{fig:kp3L}(h) and~\ref{fig:kp3L}(n) for WS$_2$, the conductances $G_{\mathrm{P},\downarrow}(\mathbf{k}_{\parallel})$ and $G_{\mathrm{AP},\uparrow}(\mathbf{k}_{\parallel})$ approach the order of one conductance quantum ($e^2/h$) at $\mathbf{k}_{\parallel}$ points around $\mathbf{k}_{2}$, implying transmission probabilities close to unity despite the presence of a tunnel barrier (see Fig.~S2 for the same behavior in MoSe$_2$ and WSe$_2$). In general, the transmission through a tunnel barrier decays exponentially with the barrier thickness $d$, i.e., $T(E,\mathbf{k}_{\parallel}) \propto e^{-2\kappa(E,\mathbf{k}_{\parallel}) d}$, and typically remains well below unity. Therefore, while the barrier decay rate explains the overall $\mathbf{k}_{\parallel}$-resolved pattern of suppressed and enhanced conductance, the emergence of these high-conductance channels cannot be understood from the electrode and barrier bulk properties alone.

\subsubsection{Effect of induced states resonance on transmission}

These additional high-conductance channels suggest the involvement of interfacial states. Hybridization between the bulk electrode states and the electronic states at the interface can induce gap states inside the semiconducting barrier. Such interface-induced metallic states have indeed been reported in Fe/MoS$_2$/Fe~\cite{Dolui2014}, NiFe/MoS$_2$/NiFe~\cite{WangWeiyi2015}, Co/MoS$_2$/Co~\cite{Huang2025}, and Ti$_2$C-MXene/MoS$_2$~\cite{Gan2013} heterostructures.

\begin{figure}
    \centering
\includegraphics[width=8.0cm,clip]{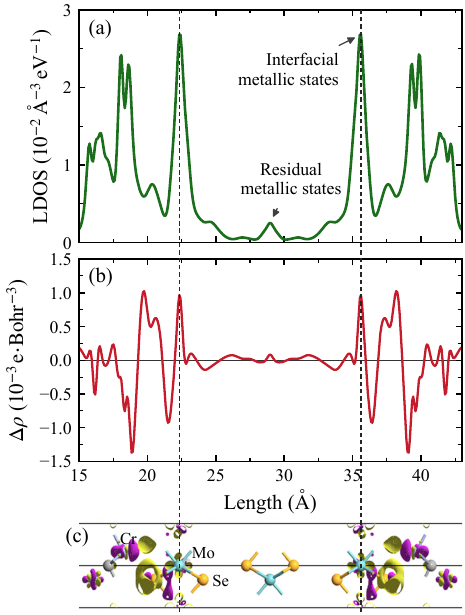}
\caption{Results for the 3L-MoSe$_2$ junction. (a) TLDOS (the sum of the majority- and minority-spin contributions in the parallel configuration) at the Fermi level along the transport direction. (b) Profile of the charge-density difference along the transport direction. (c) Isosurface of the charge-density difference (isovalue of 0.0015~$e\,\mathrm{Bohr}^{-3}$). Magenta and yellow denote charge depletion and accumulation, respectively.}
\label{fig:ch}
\end{figure}

Direct evidence for such states is provided by the \gls{ldos}. Figure~\ref{fig:ch}(a) shows the total LDOS (TLDOS, the sum of the majority- and minority-spin contributions in the parallel configuration) of the 3L-MoSe$_2$ junction at $E=E_\mathrm{F}$ along the transport direction ($c$ axis). The interface formation clearly induces metallic states in the originally insulating barrier region. These states are most pronounced at the interfacial Mo sites, consistent with previous studies~\cite{Dolui2014, Gan2013}, and gradually decay toward the center of the barrier, leaving only a small residual contribution.

The interfacial origin of these states is corroborated by the charge-redistribution maps at the Cr$_2$C/TMDC interfaces shown in Figs.~\ref{fig:ch}(b) and \ref{fig:ch}(c). Similar TLDOS features and charge redistribution are observed for MoS$_2$, WS$_2$, and WSe$_2$. In all cases, charge is depleted from the interfacial Cr atoms and accumulates on the adjacent S/Se and Mo/W atoms, indicating interfacial charge transfer and orbital hybridization at both interfaces. The hybridization is attenuated inside the barrier because of the weak \gls{vdw} interaction between adjacent barrier layers, reflected in the larger $d_{\mathrm{BB}}$ compared with $d_{\mathrm{EB}}$. Nevertheless, the overlap of the states penetrating from the left and right interfaces leaves a non-negligible metallic weight at the central Mo/W atom. Hereafter, we refer to these as \textit{interfacial states} and to their remnant in the middle of the barrier as \textit{residual states} at the barrier center. 

The connection between these interfacial states and the high-transmission channels in the WS$_2$ junction becomes evident in the $\mathbf{k}_{\parallel}$-resolved LDOS shown in Fig.~\ref{fig:kp3L}. In the following, i (c) denotes the interfacial (barrier-center) layer, and the interfacial LDOS, $N_\mathrm{i}(\mathbf{k}_{\parallel})$, is obtained by summing the left and right interfacial $Y$-$M$-$Y$ layers ($Y=$ S, Se; $M=$ Mo, W), while $N_\mathrm{c}(\mathbf{k}_{\parallel})$ denotes the LDOS of the central $Y$-$M$-$Y$ layer at $E=E_{\mathrm{F}}$. In both cases, the LDOS is concentrated away from the $\Gamma$ point. For the parallel configuration, the majority-spin interfacial LDOS, $N_{\mathrm{P,i},\uparrow}(\mathbf{k}_{\parallel})$, forms a sharp ring-shaped feature around the $\mathbf{k}_{1}$ region in MoS$_2$ [Fig.~\ref{fig:kp3L}(c)], whereas in WS$_2$ its weight shifts toward $\mathbf{k}_{2}$ [Fig.~\ref{fig:kp3L}(d)]. This difference can be attributed to the distinct interface stackings (M3 for MoS$_2$ and M1 for WS$_2$; Fig.~\ref{fig:mtj}), which modify the interfacial hybridization. In contrast, the minority-spin LDOS, $N_{\mathrm{P,i},\downarrow}(\mathbf{k}_{\parallel})$, is localized around $\mathbf{k}_{2}$ for all barriers. The majority-spin LDOS in the antiparallel configuration, $N_{\mathrm{AP,i},\uparrow}(\mathbf{k}_{\parallel})$, combines features of the two spin channels because, in this configuration, tunneling involves opposite-spin states in the two electrodes. The central-layer LDOS closely follows the $\mathbf{k}_{\parallel}$ distribution of the interfacial LDOS, albeit with reduced intensity, indicating that the residual states are not independent electronic states but arise from the overlap of the decaying tails of the interfacial states. Previous studies have demonstrated that interfacial states significantly affect the conductance when they are well coupled to the propagating states of the electrodes on both sides of the barrier~\cite{Butler2001, RUNGGER2007, Masuda2020, Masudaints2021}. If these interfacial states furthermore retain a finite metallic weight at the barrier center through overlapping tails, they can resonantly couple across the barrier, thereby increasing the transmission probability. 

This resonant-coupling picture is confirmed by comparing the $\mathbf{k}_{\parallel}$-resolved LDOSs with the corresponding conductance maps in Fig.~\ref{fig:kp3L}. The conductance closely follows both $N_\mathrm{i}(\mathbf{k}_{\parallel})$ and $N_\mathrm{c}(\mathbf{k}_{\parallel})$, although the correlation is considerably stronger for the latter [cf. Figs.~\ref{fig:kp3L}(a),~\ref{fig:kp3L}(c), and~\ref{fig:kp3L}(e) near $\mathbf{k}_{1}$]. A similar agreement between the $\mathbf{k}_{\parallel}$-integrated barrier-center LDOSs and the conductances has also been reported previously~\cite{Tanaka2023}. The central-layer LDOS makes the contrast between the two barriers explicit. In MoS$_2$, the residual states in $N_{\mathrm{P,c},\downarrow}(\mathbf{k}_{\parallel})$ [Fig.~\ref{fig:kp3L}(k)] and $N_{\mathrm{AP,c},\uparrow}(\mathbf{k}_{\parallel})$ [Fig.~\ref{fig:kp3L}(q)] are confined to a small region of the BZ, whereas in WS$_2$ they remain broadly distributed around the $\mathbf{k}_{2}$ region with significant weight [Figs.~\ref{fig:kp3L}(l) and \ref{fig:kp3L}(r)]. Correspondingly, $G_{\mathrm{P},\downarrow} (\mathbf{k}_{\parallel})$ [Fig.~\ref{fig:kp3L}(g)] and $G_{\mathrm{AP},\uparrow}(\mathbf{k}_{\parallel})$ [Fig.~\ref{fig:kp3L}(m)] are suppressed in MoS$_2$ but remain high in WS$_2$ near $\mathbf{k}_{2}$ [Figs.~\ref{fig:kp3L}(h) and \ref{fig:kp3L}(n)]. This consistency indicates that the high-transmission channels near $\mathbf{k}_{2}$ in WS$_2$ originate from strong resonant coupling between the interfacial states across the barrier, whereas the weaker residual states in MoS$_2$ suppress this resonance and hence the transmission. A similar analysis applies to the high-transmission channels observed in MoSe$_2$ and WSe$_2$ (see Fig.~S2 in the SM). Thus, the extent of resonance between the interfacial states, as evidenced by the residual weight at the barrier center, is responsible for the contrasting conductance behavior and ultimately the TMR of these TMDC-based MTJs.

Since interfacial states penetrate into the barrier and are accompanied by high-transmission channels, one may ask whether the TMDC layers remain tunnel barriers or instead behave as metallic spacers. Although the interfacial states are pronounced, they decay rapidly away from the interfaces (Fig.~\ref{fig:ch}). Consequently, the high-transmission channels are confined to localized $\mathbf{k}_{\parallel}$ regions associated with resonant coupling, whereas the conductance over the rest of the BZ is suppressed by 2--3 orders of magnitude. Furthermore, despite the presence of minority-spin states at the Fermi level in the Cr$_2$C electrodes, the conductance around the $\Gamma$ point is suppressed by 5--6 orders of magnitude, reflecting the large decay rate of the corresponding evanescent states in the barrier. In contrast, a metallic spacer typically exhibits conductance approaching or exceeding one conductance quantum, $e^2/h$, over a substantial portion of the $\mathbf{k}_{\parallel}$-resolved BZ, depending on the number of bands crossing the Fermi level~\cite{Sakuraba2010sp,Simalaotao_2025,LiXinlu2021}. The residual states therefore do not transform the central layer into a metallic spacer. Instead, they provide localized resonant tunneling pathways at selected $\mathbf{k}_{\parallel}$ points while preserving the tunneling character of the barrier.

\subsection{Dependence on barrier thickness}\label{Subs:III(B)}
\begin{figure}
\centering
\includegraphics[width=8.0cm,clip]{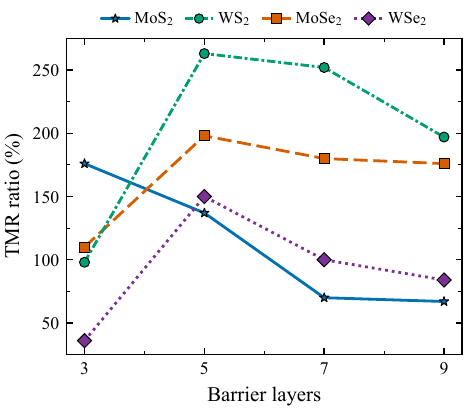}
\caption{TMR ratio as a function of barrier thickness.}
\label{fig:thick}
\end{figure}

Next, we examine the evolution of the residual states and their effect on the TMR ratio as the \gls{tmdc} barrier thickness increases from 3L to 5L, 7L, and 9L. Only odd numbers of layers are considered to preserve the structural inversion symmetry between the two interfaces. 
As shown in Fig.~\ref{fig:thick}, the TMR ratio decreases with increasing barrier thickness beyond 5L for all TMDC barriers. The reduction is most pronounced for WSe$_2$, whereas WS$_2$ and MoSe$_2$ show a comparatively gradual decrease. The behavior between 3L and 5L, on the other hand, differs qualitatively among the barriers. MoS$_2$ displays a monotonic reduction over the entire thickness range, whereas WS$_2$, WSe$_2$, and MoSe$_2$ show an increase in the TMR ratio from 3L to 5L.

\begin{figure}
    \centering
\includegraphics[width=8.0cm,clip]{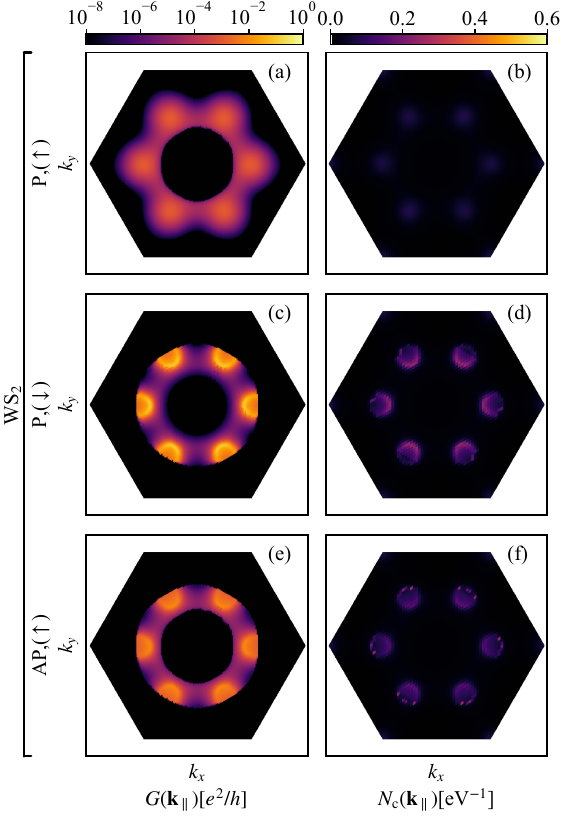}
\caption{$\mathbf{k}_{\parallel}$-resolved conductances $G(\mathbf{k}_{\parallel})$ [(a),(c),(e)] and corresponding LDOSs of central layer $N_\mathrm{c}(\mathbf{k}_{\parallel})$ [(b),(d),(f)] at $E=E_\mathrm{F}$ of 5L WS$_2$ barrier. The top row [(a),(b)] shows the majority-spin and the middle row [(c),(d)] the minority-spin in the parallel (P) magnetization configuration. The bottom row [(e),(f)] shows the majority-spin in the antiparallel (AP) magnetization configuration.}
\label{fig:kpfive}
\end{figure}

The overall decrease beyond 5L directly reflects the gradual attenuation of the residual states. Figure~\ref{fig:kpfive} shows the $\mathbf{k}_{\parallel}$-resolved conductance and the central-layer LDOS for the 5L WS$_2$ barrier. The corresponding results for the other barriers are presented in Fig.~S4 of the SM. Compared with the 3L barrier, the central-layer LDOS is significantly reduced in the 5L barrier, indicating further suppression of the residual states. Accordingly, the high-transmission channels in $G_{\mathrm{P},\downarrow}(\mathbf{k}_{\parallel})$ [Fig.~\ref{fig:kpfive}(c)] and $G_{\mathrm{AP},\uparrow}(\mathbf{k}_{\parallel})$ [Fig.~\ref{fig:kpfive}(e)] are strongly weakened relative to the 3L case [cf. Figs.~\ref{fig:kp3L}(h) and~\ref{fig:kp3L}(n)], showing that the resonance between the interfacial states across the barrier weakens with increasing thickness. For the 7L and 9L barriers, both the central-layer LDOS and the conductances are suppressed even more strongly. As a result, the resonant enhancement of the parallel conductance is progressively lost, and the reduced contrast between the parallel and antiparallel conductances leads to the decrease in the TMR ratio beyond 5L.

The initial increase in the TMR ratio from 3L to 5L for WS$_2$, MoSe$_2$, and WSe$_2$, on the other hand, has a different origin. Although the LDOSs at the barrier center are significantly suppressed at 5L, weak induced states still persist in the minority-spin components of these three barriers, as shown in Fig.~\ref{fig:kpfive}(d) for WS$_2$. The persistence of these states can be traced back to the trilayer junctions, where $N_{\mathrm{P,c},\downarrow}(\mathbf{k}_{\parallel})$ was found to be considerably larger for these three barriers than for MoS$_2$. The persisting states sustain a relatively large $G_{\mathrm{P},\downarrow}(\mathbf{k}_{\parallel})$ [Fig.~\ref{fig:kpfive}(c)], whereas $G_{\mathrm{AP},\uparrow}(\mathbf{k}_{\parallel})$ [Fig.~\ref{fig:kpfive}(e)] is suppressed more strongly. This imbalance between the two conductances gives rise to the increase in the TMR ratio from 3L to 5L.

Overall, these results show that the evolution of the induced states across the barrier dominates the transport for thinner barriers, making the optimization of the barrier thickness essential for maximizing the TMR ratio. A 5L barrier provides the optimum balance between spin-selective tunneling and barrier thickness for WS$_2$-, MoSe$_2$-, and WSe$_2$-based junctions, whereas for MoS$_2$ increasing the thickness progressively suppresses the transport channels responsible for a high TMR ratio. A similar electrode-dependent barrier-thickness dependence has been reported for MoS$_2$-based MTJs~\cite{Rotjanapittayakul2018, Dolui2014, Zhang2016, Huang2025}.

\subsection{Effect of biaxial tensile strain on trilayer barriers}\label{Subs:III(C)}
The results so far demonstrate that the transport in Cr$_2$C/TMDC junctions is governed by the induced states inside the barrier. Increasing the barrier thickness suppresses the residual states at the center, but it leaves the interfacial states themselves largely unaffected. An alternative strategy is thus to tune the interfacial states directly. Biaxial tensile strain is well suited for this purpose because it modifies the orbital hybridization and charge redistribution at the interface through changes in bond lengths, bond angles, and interlayer distances~\cite{Wang2025, Loong2014}. If the interfacial states support transport in only one spin channel, i.e., contribute to either $G_{\mathrm{P},\uparrow}$ or $G_{\mathrm{P},\downarrow}$ alone, they can also enhance the TMR ratio~\cite{Masuda2020, Masudaints2021}. A spin-selective response to strain is indeed expected for the present electrodes. Sun \textit{et al.} showed that under biaxial tensile strain, the spin-up density of states of Cr$_2$C MXene remains largely unchanged, whereas the spin-down density of states shifts toward higher energies~\cite{Sun2021}. Furthermore, both TMDCs and Cr$_2$C MXene remain thermodynamically stable under strains up to $\sim$10\%~\cite{Carrascoso2022, Sun2021}, rendering these junctions well suited for strain engineering. Motivated by these considerations, we investigate the effect of biaxial tensile strain on the trilayer junctions.

\begin{figure}
    \centering
\includegraphics[width=8.0cm, clip]{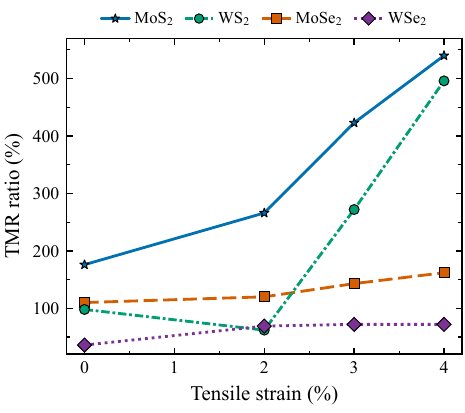}
\caption{TMR ratio as a function of biaxial tensile strain for the trilayer barriers.}
\label{fig:strain}
\end{figure}

Figure~\ref{fig:strain} shows the evolution of the TMR ratio under biaxial tensile strain. For the MoS$_2$ and WS$_2$ barriers, the TMR ratio increases sharply from 176\% and 98\% in the unstrained case to 540\% and 496\%, respectively, at 4\% strain (the increase is non-monotonic for WS$_2$, whose TMR ratio first decreases slightly at 2\% strain). In contrast, MoSe$_2$ and WSe$_2$ exhibit only a modest enhancement, with the TMR ratio increasing from 110\% to 162\% and from 36\% to 72\%, respectively.
Notably, the material dependence of the strain response is different from that of the thickness dependence. Strain is most effective for the sulfide barriers, MoS$_2$ and WS$_2$, whereas the thickness increase to 5L was most beneficial for WS$_2$, MoSe$_2$, and WSe$_2$.

\begin{figure}
    \centering
\includegraphics[width=8.0cm,clip]{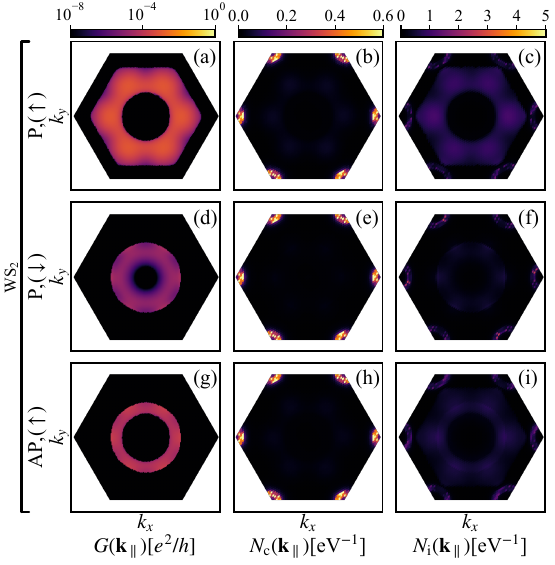}
\caption{$\mathbf{k}_{\parallel}$-resolved conductances $G(\mathbf{k}_{\parallel})$ [(a),(d),(g)] for 3L WS$_2$ barrier at $E=E_\mathrm{F}$ and 4\% tensile strain. The corresponding central layer LDOSs $N_\mathrm{c}(\mathbf{k}_{\parallel})$ are shown in column [(b),(e),(h)] and interfacial layers LDOSs $N_\mathrm{i}(\mathbf{k}_{\parallel})$ in [(c),(f),(i)]. The top row [(a)--(c)] shows the majority-spin and the middle row [(d)--(f)] the minority-spin in the parallel (P) magnetization configuration. The bottom row [(g)--(i)] shows the majority-spin in the antiparallel (AP) magnetization configuration.}
\label{fig:ldosstrain}
\end{figure}

The microscopic origin of this contrasting strain response can again be traced to the interfacial states. Figure~\ref{fig:ldosstrain} shows the $\mathbf{k}_{\parallel}$-resolved conductances of the WS$_2$ junction at 4\% strain, together with the corresponding central-layer and interfacial LDOSs. The results for the other systems are presented in Fig.~S5 of the SM. Under strain, the conductances of WS$_2$ are substantially reduced, particularly $G_{\mathrm{P},\downarrow}(\mathbf{k}_{\parallel})$ [Fig.~\ref{fig:ldosstrain}(d)] and $G_{\mathrm{AP},\uparrow}(\mathbf{k}_{\parallel})$ [Fig.~\ref{fig:ldosstrain}(g)], and the high-transmission channels observed in the unstrained junction are almost completely absent [cf. Figs.~\ref{fig:kp3L}(h) and~\ref{fig:kp3L}(n)].

These changes in the conductance reflect the strain-induced modulation of the interfacial states. As shown in Figs.~\ref{fig:ldosstrain}(b),~\ref{fig:ldosstrain}(e), and~\ref{fig:ldosstrain}(h), the central-layer LDOSs of WS$_2$ are almost completely suppressed compared with the unstrained case [Figs.~\ref{fig:kp3L}(f),~\ref{fig:kp3L}(l), and~\ref{fig:kp3L}(r)]. Because these residual states arise from the overlapping tails of the interfacial states, we next examine the interfacial LDOSs [Figs.~\ref{fig:ldosstrain}(c),~\ref{fig:ldosstrain}(f), and~\ref{fig:ldosstrain}(i)] to identify the origin of this suppression. Compared with the unstrained junction [Figs.~\ref{fig:kp3L}(d), \ref{fig:kp3L}(j), and \ref{fig:kp3L}(p)], the minority-spin component of WS$_2$ is markedly reduced by strain, whereas a considerable majority-spin component remains. The strain-induced suppression is thus spin selective. The weakened minority-spin interfacial states penetrate less deeply into the barrier, which accounts for the suppressed $N_{\mathrm{P,c},\downarrow}(\mathbf{k}_{\parallel})$ and the disappearance of the corresponding high-transmission channels around the $\mathbf{k}_{2}$ region. In the parallel configuration, the remaining majority-spin states continue to support $G_{\mathrm{P},\uparrow}$. In the antiparallel configuration, by contrast, transmission requires coupling between majority-spin states at one interface and minority-spin states at the other, so the suppression of the minority-spin component acts as a bottleneck. Consequently, $G_{\mathrm{AP}}$ decreases more strongly than $G_{\mathrm{P}}$ (Table~S2), sharply increasing the TMR ratio of WS$_2$. The same mechanism explains the strain-induced enhancement in MoS$_2$.

By contrast, WSe$_2$ [Fig.~S5] exhibits only a moderate reduction in conductance, with appreciable high-transmission channels remaining in both $G_{\mathrm{P},\downarrow}(\mathbf{k}_{\parallel})$ [Fig.~S5(m)] and $G_{\mathrm{AP},\uparrow}(\mathbf{k}_{\parallel})$ [Fig.~S5(p)]. Comparison of the LDOS with the unstrained case [Fig.~S2] shows that strain only weakly modifies the interfacial states in WSe$_2$, allowing a sizable residual weight to persist at the barrier center. Resonant coupling across the barrier is therefore largely preserved and sustains transmission in both magnetic configurations. As a result, the contrast between $G_{\mathrm{P}}$ and $G_{\mathrm{AP}}$ remains relatively small and the TMR ratio increases only modestly (Table~S2). A similar mechanism operates in the MoSe$_2$ junction.

Taken together, these results indicate that biaxial tensile strain can selectively tune the interfacial states of a particular spin channel and thereby enhance the TMR ratio. In the present systems, strain is particularly effective for the MoS$_2$- and WS$_2$-based junctions, where it suppresses the residual weight, whereas increasing the barrier thickness to 5L is the better route for the MoSe$_2$- and WSe$_2$-based junctions. These findings demonstrate that induced-state engineering via strain and barrier thickness provides complementary approaches to improving the TMR ratio in TMDC-based MTJs.

\section{Summary}

In summary, we have investigated the spin-dependent transport in all-2D \gls{vdw} MTJs of the structure Cr$_2$C/$MY_2$/Cr$_2$C ($M$ = Mo, W; $Y$ = S, Se) with barrier thicknesses of 3, 5, 7, and 9 layers using first-principles transport calculations. While the broad features of the $\mathbf{k}_{\parallel}$-resolved conductances follow the decay of the evanescent states in the barrier, the trilayer WS$_2$, MoSe$_2$, and WSe$_2$ junctions host near-unity transmission channels at the off-$\Gamma$ hot spots. These channels arise from the resonant coupling between the interfacial states at the two electrode--barrier interfaces, as evidenced by their residual weight at the barrier center, and the strength of this coupling accounts for the barrier-material dependence of the conductances and TMR ratios. The same mechanism explains the thickness dependence. As this coupling weakens in thicker barriers, the TMR ratio decreases monotonically for MoS$_2$, whereas it is maximized at five layers for the other junctions. Finally, we showed that biaxial tensile strain selectively suppresses the minority-spin interfacial states and residual weight, raising the TMR ratio from 176\% to 540\% for MoS$_2$ and from 98\% to 496\% for WS$_2$ at 4\% strain, whereas increasing the barrier thickness is the more effective route for MoSe$_2$ and WSe$_2$. These results establish interfacial-state engineering through strain and barrier thickness as complementary design strategies for TMDC-based MTJs. 

\begin{acknowledgments}
This study was partially supported by JSPS KAKENHI Grant Numbers 23H00183, 23K03933, and 25H00743, and by NIMS ICGP. The high-performance computing resources provided by NIMS are gratefully acknowledged. We thank E.~Xiao, I.~Kurniawan, and K.~Simalaotao for fruitful discussions. We also thank A.~Sharma and A.~Shah for valuable discussions.
\end{acknowledgments}

\bibliography{references}

@article{Zhangcr2cmulti2026,
  title = {{Two-dimensional Dirac half-metallic $\mathrm{Cr_2C}$ sheets with a clean surface and high stability}},
  ISSN = {2731-0582},
  url = {http://dx.doi.org/10.1038/s44160-026-01097-2},
  DOI = {10.1038/s44160-026-01097-2},
  journal = {Nature Synthesis},
  publisher = {Springer Science and Business Media LLC},
  author = {Zhang,  Yanan and Yuan,  Sheng and Gong,  Wenbin and Ju,  Tao and Li,  Pengfei and Yue,  Di and Wang,  Yunong and Shi,  Zhenzhong and Zhao,  Zhigang and Geng,  Jianxin and Zhang,  Yuanbo and Geng,  Fengxia and Cheng,  Hui-Ming},
  year = {2026},
}

@article{Masudaints2021,
  title = {Interfacial giant tunnel magnetoresistance and bulk-induced large perpendicular magnetic anisotropy in (111)-oriented junctions with fcc ferromagnetic alloys: {A} first-principles study},
  author = {Masuda, Keisuke and Itoh, Hiroyoshi and Sonobe, Yoshiaki and Sukegawa, Hiroaki and Mitani, Seiji and Miura, Yoshio},
  journal = {Phys. Rev. B},
  volume = {103},
  issue = {6},
  pages = {064427},
  numpages = {10},
  year = {2021},
  month = {Feb},
  publisher = {American Physical Society},
  doi = {10.1103/PhysRevB.103.064427},
  url = {https://link.aps.org/doi/10.1103/PhysRevB.103.064427}
}

@article{Scheikespo2022,
    author = {Scheike, Thomas and Wen, Zhenchao and Sukegawa, Hiroaki and Mitani, Seiji},
    title = {Enhanced tunnel magnetoresistance in {Fe}/$\mathrm{Mg_4Al-O_x}$/{Fe}(001) magnetic tunnel junctions},
    journal = {Applied Physics Letters},
    volume = {120},
    number = {3},
    pages = {032404},
    year = {2022},
    month = {01},
    issn = {0003-6951},
    doi = {10.1063/5.0082715},
    url = {https://doi.org/10.1063/5.0082715},
}

@article{Hiroakispo2012,
  title = {Enhanced tunnel magnetoresistance in a spinel oxide barrier with cation-site disorder},
  author = {Sukegawa, Hiroaki and Miura, Yoshio and Muramoto, Shingo and Mitani, Seiji and Niizeki, Tomohiko and Ohkubo, Tadakatsu and Abe, Kazutaka and Shirai, Masafumi and Inomata, Koichiro and Hono, Kazuhiro},
  journal = {Phys. Rev. B},
  volume = {86},
  issue = {18},
  pages = {184401},
  numpages = {5},
  year = {2012},
  month = {Nov},
  publisher = {American Physical Society},
  doi = {10.1103/PhysRevB.86.184401},
  url = {https://link.aps.org/doi/10.1103/PhysRevB.86.184401}
}

@article{Miuraspo2012,
  title = {First-principles study of tunneling magnetoresistance in {Fe}/$\mathrm{MgAl_2O_4}$/{Fe}(001) magnetic tunnel junctions},
  author = {Miura, Yoshio and Muramoto, Shingo and Abe, Kazutaka and Shirai, Masafumi},
  journal = {Phys. Rev. B},
  volume = {86},
  issue = {2},
  pages = {024426},
  numpages = {6},
  year = {2012},
  month = {Jul},
  publisher = {American Physical Society},
  doi = {10.1103/PhysRevB.86.024426},
  url = {https://link.aps.org/doi/10.1103/PhysRevB.86.024426}
}

@article{Zhang2016,
    author = {Zhang, Han and Ye, Meng and Wang, Yangyang and Quhe, Ruge and Pan, Yuanyuan and Guo, Ying and Song, Zhigang and Yang, Jinbo and Guo, Wanlin and Lu, Jing},
    title = {{Magnetoresistance in Co/2D $\mathrm{MoS_2}$/Co and Ni/2D MoS2/Ni junctions}},
    journal = {Physical Chemistry Chemical Physics},
    volume = {18},
    number = {24},
    pages = {16367-16376},
    year = {2016},
    month = {06},
    issn = {1463-9076},
    doi = {10.1039/c6cp01866a},
    url = {https://doi.org/10.1039/c6cp01866a},
}

@ARTICLE{Wu2015,
  title     = {Spin-dependent transport properties of $\mathrm{Fe_3O_4}$/$\mathrm{MoS_2}$/$\mathrm{Fe_3O_4}$
               junctions},
  author    = {Wu, Han-Chun and Coile{\'a}in, Cormac {\'O} and Abid, Mourad and
               Mauit, Ozhet and Syrlybekov, Askar and Khalid, Abbas and Xu,
               Hongjun and Gatensby, Riley and Jing Wang, Jing and Liu, Huajun
               and Yang, Li and Duesberg, Georg S and Zhang, Hong-Zhou and
               Abid, Mohamed and Shvets, Igor V},
  journal   = {Sci. Rep.},
  publisher = {Springer Science and Business Media LLC},
  volume    =  {5},
  pages     = {15984},
  year      =  {2015},
  doi = {https://doi.org/10.1038/srep15984}
}

@article{RUNGGER2007,
title = {Electronic transport through {Fe}/{MgO}/{Fe}(100) tunnel junctions},
journal = {Journal of Magnetism and Magnetic Materials},
volume = {316},
number = {2},
pages = {481-483},
year = {2007},
note = {Proceedings of the Joint European Magnetic Symposia},
issn = {0304-8853},
doi = {https://doi.org/10.1016/j.jmmm.2007.03.146},
url = {https://www.sciencedirect.com/science/article/pii/S0304885307004994},
author = {Ivan Rungger and Alexandre {Reily Rocha} and Oleg Mryasov and Olle Heinonen and Stefano Sanvito},
}

@article{LiXinlu2021,
  title = {Current-Perpendicular-to-Plane Giant Magnetoresistance Effect in van der Waals Heterostructures},
  author = {Li, Xinlu and Su, Yurong and Zhu, Meng and Zheng, Fanxing and Zhang, Peina and Zhang, Jia and L\"u, Jing-Tao},
  journal = {Phys. Rev. Appl.},
  volume = {16},
  issue = {3},
  pages = {034052},
  numpages = {10},
  year = {2021},
  month = {Sep},
  publisher = {American Physical Society},
  doi = {10.1103/PhysRevApplied.16.034052},
  url = {https://link.aps.org/doi/10.1103/PhysRevApplied.16.034052}
}

@article{Sakuraba2006,
    author = {Sakuraba, Y. and Hattori, M. and Oogane, M. and Ando, Y. and Kato, H. and Sakuma, A. and Miyazaki, T. and Kubota, H.},
    title = {Giant tunneling magnetoresistance in $\mathrm{Co_2MnSi}$/{Al-O}/$\mathrm{Co_2MnSi}$ magnetic tunnel junctions},
    journal = {Applied Physics Letters},
    volume = {88},
    number = {19},
    pages = {192508},
    year = {2006},
    month = {05},
    issn = {0003-6951},
    doi = {10.1063/1.2202724},
    url = {https://doi.org/10.1063/1.2202724},
}

@article{DavidD2005,
    author = {Djayaprawira, David D. and Tsunekawa, Koji and Nagai, Motonobu and Maehara, Hiroki and Yamagata, Shinji and Watanabe, Naoki and Yuasa, Shinji and Suzuki, Yoshishige and Ando, Koji},
    title = {230\% room-temperature magnetoresistance in {CoFeB}/{MgO}/{CoFeB} magnetic tunnel junctions},
    journal = {Applied Physics Letters},
    volume = {86},
    number = {9},
    pages = {092502},
    year = {2005},
    month = {02},
    issn = {0003-6951},
    doi = {10.1063/1.1871344},
    url = {https://doi.org/10.1063/1.1871344},
}

@article{LeeM2006,
    author = {Lee, Y. M. and Hayakawa, J. and Ikeda, S. and Matsukura, F. and Ohno, H.},
    title = {Giant tunnel magnetoresistance and high annealing stability in {CoFeB}/{MgO}/{CoFeB} magnetic tunnel junctions with synthetic pinned layer},
    journal = {Applied Physics Letters},
    volume = {89},
    number = {4},
    pages = {042506},
    year = {2006},
    month = {07},
    issn = {0003-6951},
    doi = {10.1063/1.2234720},
    url = {https://doi.org/10.1063/1.2234720},
}

@article{Miura2008,
  title = {Half-metallic interface and coherent tunneling in $\mathrm{Co_{2}YZ}$/{MgO}/$\mathrm{Co_{2}YZ}$ ({YZ}={MnSi},{CrAl}) magnetic tunnel junctions: {A} first-principles study},
  author = {Miura, Yoshio and Uchida, Hirohisa and Oba, Yoshihiro and Abe, Kazutaka and Shirai, Masafumi},
  journal = {Phys. Rev. B},
  volume = {78},
  issue = {6},
  pages = {064416},
  numpages = {9},
  year = {2008},
  month = {Aug},
  publisher = {American Physical Society},
  doi = {10.1103/PhysRevB.78.064416},
  url = {https://link.aps.org/doi/10.1103/PhysRevB.78.064416}
}

@article{Borge2023,
author = {Borge-Dur{\'a}n, Ignacio and Paul, Atanu and Grinberg, Ilya},
title = {{From Non-Magnetic to Magnetic: A First-Principles Study of the Emergence of Magnetism in 2D $\mathrm{(Nb_{1-x}Ti_{x})_4C_3}$ MXenes}},
journal = {Chemistry of Materials},
volume = {35},
number = {18},
pages = {7442-7449},
year = {2023},
doi = {10.1021/acs.chemmater.3c00367},
URL = {https://doi.org/10.1021/acs.chemmater.3c00367},
}

@article{Aull2022,
  title = {{Ab Initio Study of Magnetic Tunnel Junctions Based on Half-Metallic and Spin-Gapless Semiconducting Heusler Compounds: Reconfigurable Diode and Inverse Tunnel-Magnetoresistance Effect}},
  author = {Aull, T. and \ifmmode \mbox{\c{S}}\else \c{S}\fi{}a\ifmmode \mbox{\c{s}}\else \c{s}\fi{}\ifmmode \imath \else \i \fi{}o\ifmmode \breve{g}\else \u{g}\fi{}lu, E. and Hinsche, N.F. and Mertig, I.},
  journal = {Phys. Rev. Appl.},
  volume = {18},
  issue = {3},
  pages = {034024},
  numpages = {14},
  year = {2022},
  month = {Sep},
  publisher = {American Physical Society},
  doi = {10.1103/PhysRevApplied.18.034024},
  url = {https://link.aps.org/doi/10.1103/PhysRevApplied.18.034024}
}

@article{Sergey2017,
  title = {Heusler compounds with perpendicular magnetic anisotropy and large tunneling magnetoresistance},
  author = {Faleev, Sergey V. and Ferrante, Yari and Jeong, Jaewoo and Samant, Mahesh G. and Jones, Barbara and Parkin, Stuart S. P.},
  journal = {Phys. Rev. Mater.},
  volume = {1},
  issue = {2},
  pages = {024402},
  numpages = {9},
  year = {2017},
  month = {Jul},
  publisher = {American Physical Society},
  doi = {10.1103/PhysRevMaterials.1.024402},
  url = {https://link.aps.org/doi/10.1103/PhysRevMaterials.1.024402}
}

@article{Sakuraba2010sp,
  title = {Mechanism of large magnetoresistance in $\mathrm{Co_2MnSi}$/\text{Ag}/$\mathrm{Co_2MnSi}$ devices with current perpendicular to the plane},
  author = {Sakuraba, Y. and Izumi, K. and Iwase, T. and Bosu, S. and Saito, K. and Takanashi, K. and Miura, Y. and Futatsukawa, K. and Abe, K. and Shirai, M.},
  journal = {Phys. Rev. B},
  volume = {82},
  issue = {9},
  pages = {094444},
  numpages = {5},
  year = {2010},
  month = {Sep},
  publisher = {American Physical Society},
  doi = {10.1103/PhysRevB.82.094444},
  url = {https://link.aps.org/doi/10.1103/PhysRevB.82.094444}
}

@article{Simalaotao_2025,
doi = {10.1088/1361-6463/ae06ad},
url = {https://doi.org/10.1088/1361-6463/ae06ad},
year = {2025},
month = {sep},
publisher = {IOP Publishing},
volume = {58},
number = {39},
pages = {395303},
author = {Simalaotao, Kodchakorn and Kurniawan, Ivan and Miura, Yoshio and Sakuraba, Yuya},
title = {Searching for {Cu}-{X} spacers using a half-metallic $\mathrm{Co_2FeGa_{0.5}Ge_{0.5}}$ electrode to boost magnetoresistance in {CPP}-{GMR} devices: a first-principles study},
journal = {Journal of Physics D: Applied Physics},
}

@article{Lishu2021,
    author = {Zhang, Lishu and Zhou, Jun and Li, Hui and Shen, Lei and Feng, Yuan Ping},
    title = {Recent progress and challenges in magnetic tunnel junctions with {2D} materials for spintronic applications},
    journal = {Applied Physics Reviews},
    volume = {8},
    number = {2},
    pages = {021308},
    year = {2021},
    month = {04},
    issn = {1931-9401},
    doi = {10.1063/5.0032538},
    url = {https://doi.org/10.1063/5.0032538},
}

@article{Oleg2009,
  title = {Magnetoresistive junctions based on epitaxial graphene and hexagonal boron nitride},
  author = {Yazyev, Oleg V. and Pasquarello, Alfredo},
  journal = {Phys. Rev. B},
  volume = {80},
  issue = {3},
  pages = {035408},
  numpages = {5},
  year = {2009},
  month = {Jul},
  publisher = {American Physical Society},
  doi = {10.1103/PhysRevB.80.035408},
  url = {https://link.aps.org/doi/10.1103/PhysRevB.80.035408},
}

@article{Piquemal2018,
author = {Piquemal-Banci, Ma{\"e}lis and Galceran, Regina and Godel, Florian and Caneva, Sabina and Martin, Marie-Blandine and Weatherup, Robert S. and Kidambi, Piran R. and Bouzehouane, Karim and Xavier, Stephane and Anane, Abdelmadjid and Petroff, Fr{\'e}d{\'e}ric and Fert, Albert and Dubois, Simon Mutien-Marie and Charlier, Jean-Christophe and Robertson, John and Hofmann, Stephan and Dlubak, Bruno and Seneor, Pierre},
title = {Insulator-to-{Metallic} {Spin}-{Filtering} in {2D}-{Magnetic} {Tunnel} {Junctions} {B}ased on {H}exagonal {B}oron {N}itride},
journal = {ACS Nano},
volume = {12},
number = {5},
pages = {4712-4718},
year = {2018},
doi = {10.1021/acsnano.8b01354},
URL = {https://doi.org/10.1021/acsnano.8b01354},
}

@article{Gokaran2022,
  title = {Fe- and {Co}-based magnetic tunnel junctions with {AlN} and {ZnO} spacers},
  author = {Shukla, Gokaran and Sanvito, Stefano and Lee, Geunsik},
  journal = {Phys. Rev. B},
  volume = {105},
  issue = {18},
  pages = {184427},
  numpages = {8},
  year = {2022},
  month = {May},
  publisher = {American Physical Society},
  doi = {10.1103/PhysRevB.105.184427},
  url = {https://link.aps.org/doi/10.1103/PhysRevB.105.184427},
}

@article{Ram2025,
    author = {Abhisri, Arya and Krishna Ghosh, Ram},
    title = {Resonant tunneling driven spin torque enhancement in magnetic tunnel junctions: {A} {DFT}-{NEGF} simulation study},
    journal = {Journal of Applied Physics},
    volume = {138},
    number = {12},
    pages = {123901},
    year = {2025},
    month = {09},
    issn = {0021-8979},
    doi = {10.1063/5.0285986},
    url = {https://doi.org/10.1063/5.0285986},
}

@article{Ikeda2008,
    author = {Ikeda, S. and Hayakawa, J. and Ashizawa, Y. and Lee, Y. M. and Miura, K. and Hasegawa, H. and Tsunoda, M. and Matsukura, F. and Ohno, H.},
    title = {Tunnel magnetoresistance of 604\% at 300{K} by suppression of {Ta} diffusion in {CoFeB}/{MgO}/{CoFeB} pseudo-spin-valves annealed at high temperature},
    journal = {Applied Physics Letters},
    volume = {93},
    number = {8},
    pages = {082508},
    year = {2008},
    month = {08},
    issn = {0003-6951},
    doi = {10.1063/1.2976435},
    url = {https://doi.org/10.1063/1.2976435},
}

@article{Sukegawa2010,
    author = {Sukegawa, Hiroaki and Xiu, Huixin and Ohkubo, Tadakatsu and Furubayashi, Takao and Niizeki, Tomohiko and Wang, Wenhong and Kasai, Shinya and Mitani, Seiji and Inomata, Koichiro and Hono, Kazuhiro},
    title = {Tunnel magnetoresistance with improved bias voltage dependence in lattice-matched {Fe}/spinel $\mathrm{MgAl_2O_4}$/{Fe}(001) junctions},
    journal = {Applied Physics Letters},
    volume = {96},
    number = {21},
    pages = {212505},
    year = {2010},
    month = {05},
    issn = {0003-6951},
    doi = {10.1063/1.3441409},
    url = {https://doi.org/10.1063/1.3441409},
}

@article{Scheike2023,
    author = {Scheike, Thomas and Wen, Zhenchao and Sukegawa, Hiroaki and Mitani, Seiji},
    title = {631\% room temperature tunnel magnetoresistance with large oscillation effect in {CoFe/MgO/CoFe}(001) junctions},
    journal = {Applied Physics Letters},
    volume = {122},
    number = {11},
    pages = {112404},
    year = {2023},
    month = {03},
    doi = {10.1063/5.0145873},
    url = {https://doi.org/10.1063/5.0145873},
}

@article{Sukegawa2017,
    author = {Sukegawa, Hiroaki and Kato, Yushi and Belmoubarik, Mohamed and Cheng, P.-H. and Daibou, Tadaomi and Shimomura, Naoharu and Kamiguchi, Yuuzo and Ito, Junichi and Yoda, Hiroaki and Ohkubo, Tadakatsu and Mitani, Seiji and Hono, Kazuhiro},
    title = {$\mathrm{MgGa_2O_4}$ spinel barrier for magnetic tunnel junctions: Coherent tunneling and low barrier height},
    journal = {Applied Physics Letters},
    volume = {110},
    number = {12},
    pages = {122404},
    year = {2017},
    month = {03},
    issn = {0003-6951},
    doi = {10.1063/1.4977946},
    url = {https://doi.org/10.1063/1.4977946},
}

@article{Julliere1975,
   author = {M. Julli{\`e}re},
   title = {Tunneling between ferromagnetic films},
   journal = {Phys. Lett. A},
   volume = {54},
   pages = {225-226},
   issue = {3},
   year = {1975},
}

@article{Grimme2010,
   author = {Grimme, Stefan and Antony, Jens and Ehrlich, Stephan and Krieg, Helge},
   title = {A consistent and accurate ab initio parametrization of density functional dispersion correction ({DFT-D}) for the 94 elements {H-Pu}},
   journal = {J. Chem. Phys.},
   volume = {132},
   pages = {154104},
   year = {2010},
   doi = {10.1063/1.3382344},
}

@article{Schlipf2015,
   author = {Schlipf, Martin and Gygi, Fran{\c c}ois},
   title = {Optimization algorithm for the generation of {ONCV} pseudopotentials},
   journal = {Comput. Phys. Commun.},
   volume = {196},
   pages = {36--44},
   year = {2015},
   doi = {10.1016/j.cpc.2015.05.011},
}

@article{Mathon2001,
  title = {Theory of tunneling magnetoresistance of an epitaxial $\mathrm{Fe}|\mathrm{MgO}|\mathrm{Fe}$(001) junction},
  author = {Mathon, J. and Umerski, A.},
  journal = {Phys. Rev. B},
  volume = {63},
  issue = {22},
  pages = {220403},
  numpages = {4},
  year = {2001},
  month = {May},
  publisher = {American Physical Society},
  doi = {10.1103/PhysRevB.63.220403},
  url = {https://link.aps.org/doi/10.1103/PhysRevB.63.220403},
}

@article{Butler2001,
  title = {Spin-dependent tunneling conductance of $\mathrm{Fe}|\mathrm{MgO}|\mathrm{Fe}$ sandwiches},
  author = {Butler, W. H. and Zhang, X.-G. and Schulthess, T. C. and MacLaren, J. M.},
  journal = {Phys. Rev. B},
  volume = {63},
  issue = {5},
  pages = {054416},
  numpages = {12},
  year = {2001},
  month = {Jan},
  publisher = {American Physical Society},
  doi = {10.1103/PhysRevB.63.054416},
  url = {https://link.aps.org/doi/10.1103/PhysRevB.63.054416},
}

@article{Yuasa2004,
  title = {Giant room-temperature magnetoresistance in single-crystal $\mathrm{Fe}|\mathrm{MgO}|\mathrm{Fe}$ magnetic tunnel junctions},
  volume = {3},
  ISSN = {1476-4660},
  url = {http://dx.doi.org/10.1038/nmat1257},
  DOI = {10.1038/nmat1257},
  number = {12},
  journal = {Nature Materials},
  publisher = {Springer Science and Business Media LLC},
  author = {Yuasa, Shinji and Nagahama, Taro and Fukushima, Akio and Suzuki, Yoshishige and Ando,  Koji},
  year = {2004},
  month = Oct,
  pages = {868--871},
}

@article{Parkin2004,
  title = {Giant tunnelling magnetoresistance at room temperature with $\mathrm{MgO}$(100) tunnel barriers},
  volume = {3},
  ISSN = {1476-4660},
  url = {http://dx.doi.org/10.1038/nmat1256},
  DOI = {10.1038/nmat1256},
  number = {12},
  journal = {Nature Materials},
  publisher = {Springer Science and Business Media LLC},
  author = {Parkin,  Stuart S. P. and Kaiser,  Christian and Panchula,  Alex and Rice,  Philip M. and Hughes,  Brian and Samant,  Mahesh and Yang,  See-Hun},
  year = {2004},
  month = Oct,
  pages = {862--867},
}

@article{Deng2018,
  title = {Stability of direct band gap under mechanical strains for monolayer $\mathrm{MoS_2}$,  $\mathrm{MoSe_2}$,  $\mathrm{WS_2}$ and $\mathrm{WSe_2}$},
  volume = {101},
  ISSN = {1386-9477},
  url = {http://dx.doi.org/10.1016/j.physe.2018.03.016},
  DOI = {10.1016/j.physe.2018.03.016},
  journal = {Physica E: Low-dimensional Systems and Nanostructures},
  publisher = {Elsevier BV},
  author = {Deng,  Shuo and Li,  Lijie and Li,  Min},
  year = {2018},
  month = Jul,
  pages = {44--49},
}

@article{Brandbyge2002,
   title = {Density-functional method for nonequilibrium electron transport},
  volume = {65},
  ISSN = {1095-3795},
  url = {http://dx.doi.org/10.1103/PhysRevB.65.165401},
  DOI = {10.1103/physrevb.65.165401},
  number = {16},
  pages = {165401},
  journal = {Physical Review B},
  publisher = {American Physical Society (APS)},
  author = {Brandbyge, Mads and Mozos, Jos{\'e}-Luis and Ordej{\'o}n, Pablo and Taylor, Jeremy and Stokbro,  Kurt},
  year = {2002},
  month = Mar, 
}

@article{Gan2013,
   title = {First-principles analysis of $\mathrm{MoS_{2}}$/$\mathrm{Ti_{2}C}$ and $\mathrm{MoS_{2}}$/$\mathrm{Ti_{2}C{Y}_{2}}$ ($\mathrm{Y=F, OH}$) all-{2D} semiconductor/metal contacts},
  author = {Gan, Li-Yong and Zhao, Yu-Jun and Huang, Dan and Schwingenschl\"ogl, Udo},
  journal = {Phys. Rev. B},
  volume = {87},
  issue = {24},
  pages = {245307},
  numpages = {7},
  year = {2013},
  month = {Jun},
  publisher = {American Physical Society},
  doi = {10.1103/PhysRevB.87.245307},
  url = {https://link.aps.org/doi/10.1103/PhysRevB.87.245307},
}

@article{Dolui2014,
    title = {Efficient spin injection and giant magnetoresistance in $\mathrm{Fe}|\mathrm{MoS_2}|\mathrm{Fe}$ junctions},
  author = {Dolui, Kapildeb and Narayan, Awadhesh and Rungger, Ivan and Sanvito, Stefano},
  journal = {Phys. Rev. B},
  volume = {90},
  issue = {4},
  pages = {041401},
  numpages = {5},
  year = {2014},
  month = {Jul},
  publisher = {American Physical Society},
  doi = {10.1103/PhysRevB.90.041401},
  url = {https://link.aps.org/doi/10.1103/PhysRevB.90.041401},
}

@article{Loong2014,
   title = {Strain-enhanced tunneling magnetoresistance in $\mathrm{MgO}$ magnetic tunnel junctions},
  volume = {4},
  pages = {6505},
  ISSN = {2045-2322},
  url = {http://dx.doi.org/10.1038/srep06505},
  DOI = {10.1038/srep06505},
  number = {1},
  journal = {Scientific Reports},
  publisher = {Springer Science and Business Media LLC},
  author = {Loong,  Li Ming and Qiu,  Xuepeng and Neo,  Zhi Peng and Deorani,  Praveen and Wu,  Yang and Bhatia,  Charanjit S. and Saeys,  Mark and Yang,  Hyunsoo},
  year = {2014},
}

@article{Smidstrup2017,
   title = {First-principles Green's-function method for surface calculations: A pseudopotential localized basis set approach},
  author = {Smidstrup, S\o{}ren and Stradi, Daniele and Wellendorff, Jess and Khomyakov, Petr A. and Vej-Hansen, Ulrik G. and Lee, Maeng-Eun and Ghosh, Tushar and J\'onsson, Elvar and J\'onsson, Hannes and Stokbro, Kurt},
  journal = {Phys. Rev. B},
  volume = {96},
  issue = {19},
  pages = {195309},
  numpages = {17},
  year = {2017},
  month = {Nov},
  publisher = {American Physical Society},
  doi = {10.1103/PhysRevB.96.195309},
  url = {https://link.aps.org/doi/10.1103/PhysRevB.96.195309},
}

@article{Masuda2017,
  title = {Bias voltage effects on tunneling magnetoresistance in $\mathrm{Fe}|\mathrm{MgAl_{2}O_{4}}|\mathrm{Fe}$(001) junctions: Comparative study with $\mathrm{Fe}|\mathrm{MgO}|\mathrm{Fe}$(001) junctions},
  author = {Masuda, Keisuke and Miura, Yoshio},
  journal = {Phys. Rev. B},
  volume = {96},
  issue = {5},
  pages = {054428},
  numpages = {8},
  year = {2017},
  month = {Aug},
  publisher = {American Physical Society},
  doi = {10.1103/PhysRevB.96.054428},
  url = {https://link.aps.org/doi/10.1103/PhysRevB.96.054428},
}

@article{Rotjanapittayakul2018,
   title = {Spin injection and magnetoresistance in $\mathrm{MoS_{2}}$-based tunnel junctions using $\mathrm{Fe_3Si}$ Heusler alloy electrodes},
  volume = {8},
  pages = {4779},
  ISSN = {2045-2322},
  url = {http://dx.doi.org/10.1038/s41598-018-22910-9},
  DOI = {10.1038/s41598-018-22910-9},
  number = {1},
  journal = {Scientific Reports},
  publisher = {Springer Science and Business Media LLC},
  author = {Rotjanapittayakul,  Worasak and Pijitrojana,  Wanchai and Archer,  Thomas and Sanvito,  Stefano and Prasongkit,  Jariyanee},
  year = {2018},
  month = {Mar}, 
}

@article{Smidstrup2020,
   author = {S{\o}ren Smidstrup and Troels Markussen and Pieter Vancraeyveld and Jess Wellendorff and Julian Schneider and Tue Gunst and Brecht Verstichel and Daniele Stradi and Petr A. Khomyakov and Ulrik G. Vej-Hansen and Maeng Eun Lee and Samuel T. Chill and Filip Rasmussen and Gabriele Penazzi and Fabiano Corsetti and Ari Ojanper{\"a} and Kristian Jensen and Mattias L.N. Palsgaard and Umberto Martinez and Anders Blom and Mads Brandbyge and Kurt Stokbro},
   doi = {10.1088/1361-648X/ab4007},
   issn = {1361648X},
   issue = {1},
   journal = {Journal of Physics Condensed Matter},
   pmid = {31470430},
   publisher = {Institute of Physics Publishing},
   title = {QuantumATK: An integrated platform of electronic and atomic-scale modelling tools},
   volume = {32},
   pages = {015901},
   year = {2020},
}

@article{Piquemal-Banci2020,
   author = {Ma{\"e}lis Piquemal-Banci and Regina Galceran and Simon M.M. Dubois and Victor Zatko and Marta Galbiati and Florian Godel and Marie Blandine Martin and Robert S. Weatherup and Fr{\'e}d{\'e}ric Petroff and Albert Fert and Jean Christophe Charlier and John Robertson and Stephan Hofmann and Bruno Dlubak and Pierre Seneor},
   doi = {10.1038/s41467-020-19420-6},
   issn = {20411723},
   issue = {1},
   journal = {Nature Communications},
   month = {12},
   pmid = {33168805},
   publisher = {Nature Research},
   title = {Spin filtering by proximity effects at hybridized interfaces in spin-valves with {2D} graphene barriers},
   volume = {11},
   pages = {5670},
   year = {2020},
}

@article{Lin2020,
   author = {Hailong Lin and Faguang Yan and Ce Hu and Quanshan Lv and Wenkai Zhu and Ziao Wang and Zhongming Wei and Kai Chang and Kaiyou Wang},
   doi = {10.1021/acsami.0c12483},
   issn = {19448252},
   issue = {39},
   journal = {ACS Applied Materials and Interfaces},
   month = {9},
   pages = {43921-43926},
   pmid = {32878440},
   publisher = {American Chemical Society},
   title = {Spin-Valve Effect in $\mathrm{Fe_3GeTe_2}/\mathrm{MoS_{2}}/\mathrm{Fe_3GeTe_2}$ van der $\mathrm{{Waals}}$ Heterostructures},
   volume = {12},
   year = {2020},
}

@article{Masuda2020,
   title = {Interface-driven giant tunnel magnetoresistance in (111)-oriented junctions},
  author = {Masuda, Keisuke and Itoh, Hiroyoshi and Miura, Yoshio},
  journal = {Phys. Rev. B},
  volume = {101},
  issue = {14},
  pages = {144404},
  numpages = {5},
  year = {2020},
  month = {Apr},
  publisher = {American Physical Society},
  doi = {10.1103/PhysRevB.101.144404},
  url = {https://link.aps.org/doi/10.1103/PhysRevB.101.144404},
}

@article{Sun2021,
title = {Manipulation of electronic and magnetic properties of $\mathrm{Cr_2CX_2 (X=F,O,OH)}$ monolayer by applying mechanical strains},
journal = {Journal of Alloys and Compounds},
volume = {850},
pages = {156769},
year = {2021},
issn = {0925-8388},
doi = {https://doi.org/10.1016/j.jallcom.2020.156769},
url = {https://www.sciencedirect.com/science/article/pii/S0925838820331339},
author = {Qian Sun and Zhaoming Fu and Yi Li and Zongxian Yang},
}

@article{Das2022,
author ={Das, Shreeja and Kabiraj, Arnab and Mahapatra, Santanu},
title  ={Room temperature giant magnetoresistance in half-metallic $\mathrm{Cr_2C}$ based two-dimensional tunnel junctions},
journal  ={Nanoscale},
year  ={2022},
volume  ={14},
issue  ={26},
pages  ={9409-9418},
publisher  ={The Royal Society of Chemistry},
doi  ={10.1039/D2NR02056D},
url  ={http://dx.doi.org/10.1039/D2NR02056D},
}

@article{Carrascoso2022,
title = {Biaxial versus uniaxial strain tuning of single-layer $\mathrm{MoS_{2}}$},
journal = {Nano Materials Science},
volume = {4},
number = {1},
pages = {44-51},
year = {2022},
note = {Special issue on Graphene and 2D Alternative Materials},
issn = {2589-9651},
doi = {https://doi.org/10.1016/j.nanoms.2021.03.001},
url = {https://www.sciencedirect.com/science/article/pii/S258996512100012X},
author = {Felix Carrascoso and Riccardo Frisenda and Andres Castellanos-Gomez},
}

@article{Hirohata2022,
AUTHOR={Hirohata, Atsufumi  and Elphick, Kelvin  and Lloyd, David C.  and Mizukami, Shigemi },
TITLE={Interfacial quality to control tunnelling magnetoresistance},
JOURNAL={Frontiers in Physics},
VOLUME={10},
pages = {1007989},
YEAR={2022},
URL={https://www.frontiersin.org/journals/physics/articles/10.3389/fphy.2022.1007989},
DOI={10.3389/fphy.2022.1007989},
ISSN={2296-424X},
}

@article{Jin2023,
author = {Jin, Wen and Zhang, Gaojie and Wu, Hao and Yang, Li and Zhang, Wenfeng and Chang, Haixin},
title = {Room-Temperature and Tunable Tunneling Magnetoresistance in $\mathrm{Fe_3GaTe_2}$ -Based {2D} van der {Waals} Heterojunctions},
journal = {ACS Applied Materials \& Interfaces},
volume = {15},
number = {30},
pages = {36519-36526},
year = {2023},
doi = {10.1021/acsami.3c06167},
URL = {https://doi.org/10.1021/acsami.3c06167},
}

@article{Yu2023,
author ={Yu, Hailin and Chen, Mingyan and Shao, Zhenguang and Tao, Yongmei and Jiang, Xuefan and Dong, Yaojun and Zhang, Jie and Yang, Xifeng and Liu, Yushen},
title  ={Giant tunneling magnetoresistance in in-plane double-barrier magnetic tunnel junctions based on {MXene} $\mathrm{Cr_2C}$ },
journal  ={Phys. Chem. Chem. Phys.},
year  ={2023},
volume  ={25},
issue  ={15},
pages  ={10991-10997},
publisher  ={The Royal Society of Chemistry},
doi  ={10.1039/D3CP00303E},
url  ={http://dx.doi.org/10.1039/D3CP00303E},
}

@article{Tanaka2023,
  title = {Local density of states as a probe for tunneling magnetoresistance effect: Application to ferrimagnetic tunnel junctions},
  author = {Tanaka, Katsuhiro and Nomoto, Takuya and Arita, Ryotaro},
  journal = {Phys. Rev. B},
  volume = {107},
  issue = {21},
  pages = {214442},
  numpages = {12},
  year = {2023},
  month = {Jun},
  publisher = {American Physical Society},
  doi = {10.1103/PhysRevB.107.214442},
  url = {https://link.aps.org/doi/10.1103/PhysRevB.107.214442},
}

@article{Pan2024,
   author = {Haiyang Pan and Anil Kumar Singh and Chusheng Zhang and Xueqi Hu and Jiayu Shi and Liheng An and Naizhou Wang and Ruihuan Duan and Zheng Liu and Stuart S.P. Parkin and Pritam Deb and Weibo Gao},
   doi = {10.1002/inf2.12504},
   issn = {25673165},
   issue = {6},
   journal = {InfoMat},
   month = {6},
   publisher = {John Wiley and Sons Inc},
   title = {Room-temperature tunable tunneling magnetoresistance in $\mathrm{Fe_3GaTe_2}/\mathrm{WSe_2}/\mathrm{Fe_3GaTe_2}$ van der {Waals} heterostructures},
   volume = {6},
   year = {2024},
}

@article{Huang2025,
author = {Huang, Ting-Chun and Chen, Yu-Xin and Chen, Yu-Lin and Wu, Meng-Ting and Pai, Chi-Feng and Chuang, Chiashain and Hsieh, Ya-Ping and Hofmann, Mario},
title = {Observing and Suppressing Metallization in $\mathrm{MoS_{2}}$ for Near-Ideal Spin Filtering},
journal = {ACS Applied Materials \& Interfaces},
volume = {17},
number = {48},
pages = {65967-65975},
year = {2025},
doi = {10.1021/acsami.5c15955},
URL = {https://doi.org/10.1021/acsami.5c15955},
}

@article{Shukla2025,
   author = {Gokaran Shukla and Roshan Ali and Aamir Shafique},
   doi = {10.1021/acsaelm.4c01583},
   issn = {26376113},
   issue = {1},
   journal = {ACS Applied Electronic Materials},
   month = {1},
   pages = {115-128},
   publisher = {American Chemical Society},
   title = {{Co-TMDC MTJs} : {A New Frontier in Spintronics}},
   volume = {7},
   year = {2025},
}

@article{Kurniawan2025,
   author = {I Kurniawan and K Masuda and Y Miura},
   doi = {10.1088/1361-6463/ae1b18},
   issn = {0022-3727},
   issue = {46},
   journal = {Journal of Physics D: Applied Physics},
   month = {11},
   pages = {465302},
   publisher = {IOP Publishing},
   title = {Theory of tunnel magnetoresistance in magnetic tunnel junctions with hexagonal boron nitride barriers: mechanism and application to ferromagnetic alloy electrodes},
   volume = {58},
   year = {2025},
}

@article{Kumar2025a,
   author = {Prabhat Kumar and Shunsuke Tsuda and Koichiro Yaji and Shinji Isogami},
   doi = {10.1080/14686996.2025.2551484},
   issn = {18785514},
   issue = {1},
   journal = {Science and Technology of Advanced Materials},
   publisher = {Taylor and Francis Ltd.},
   title = {Origin of two-dimensional $\mathrm{MXene}$/ferromagnetic interface evaluated by angle-dependent hard X-ray photoemission spectroscopy},
   volume = {26},
   pages = {2551484},
   year = {2025},
}

@article{Kumar2025b,
   author = {Prabhat Kumar and Yoshio Miura and Yoshinori Kotani and Akiho Sumiyoshiya and Tetsuya Nakamura and Gaurav K. Shukla and Shinji Isogami},
   doi = {10.1002/smll.202500626},
   issn = {16136829},
   issue = {25},
   journal = {Small},
   month = {6},
   pmid = {40370265},
   publisher = {John Wiley and Sons Inc},
   title = {Unconventional Spin--Orbit Torques by {2D Multilayered MXenes} for Future Nonvolatile Magnetic Memories},
   volume = {21},
   pages = {2500626},
   year = {2025},
}

@article{Wang2025,
   author = {Xinbiao Wang and Jiao Xu and Euyheon Hwang and Ji Sang Park},
   doi = {10.1063/5.0281223},
   issn = {00036951},
   issue = {12},
   journal = {Applied Physics Letters},
   month = {9},
   publisher = {American Institute of Physics},
   title = {Strain-driven interfacial engineering in metal-$\mathrm{MoS_{2}}$ contacts},
   volume = {127},
   pages = {123104},
   year = {2025},
}

@article{SG152013,
  title = {Optimized norm-conserving Vanderbilt pseudopotentials},
  author = {Hamann, D. R.},
  journal = {Phys. Rev. B},
  volume = {88},
  issue = {8},
  pages = {085117},
  numpages = {10},
  year = {2013},
  month = {Aug},
  publisher = {American Physical Society},
  doi = {10.1103/PhysRevB.88.085117},
  url = {https://link.aps.org/doi/10.1103/PhysRevB.88.085117}
}

@article{WangWeiyi2015,
author = {Wang, Weiyi and Narayan, Awadhesh and Tang, Lei and Dolui, Kapildeb and Liu, Yanwen and Yuan, Xiang and Jin, Yibo and Wu, Yizheng and Rungger, Ivan and Sanvito, Stefano and Xiu, Faxian},
title = {Spin-Valve Effect in {NiFe}/$\mathrm{MoS_2}$/{NiFe} Junctions},
journal = {Nano Letters},
volume = {15},
number = {8},
pages = {5261-5267},
year = {2015},
doi = {10.1021/acs.nanolett.5b01553},
URL = {https://doi.org/10.1021/acs.nanolett.5b01553},
}

\end{document}

%\preprint{APS/123-QED}

\title{Supplemental Material for \\
``Effects of Interfacial States and Strain on Tunnel Magnetoresistance in van der Waals Magnetic Tunnel Junctions''}% Force line breaks with \\

\author{Sakshi Goel}
\author{Arti Kashyap}
\email{arti@iitmandi.ac.in}
\affiliation{School of Physical Sciences, Indian Institute of Technology Mandi, Himachal Pradesh, India 175075}
\author{Keisuke Masuda}
\email{Masuda.Keisuke@nims.go.jp}
\author{Terumasa Tadano}
\email{Tadano.Terumasa@nims.go.jp}
\affiliation{National Institute for Materials Science (NIMS), 1-2-1 Sengen, Tsukuba, Ibaraki 305-0047, Japan}

\maketitle

% Number figures and tables as S1, S2, ...
\setcounter{figure}{0}
\setcounter{table}{0}
\renewcommand{\thefigure}{S\arabic{figure}}
\renewcommand{\thetable}{S\arabic{table}}

\begin{figure}[h]
    \centering
    \includegraphics[width=1\linewidth]{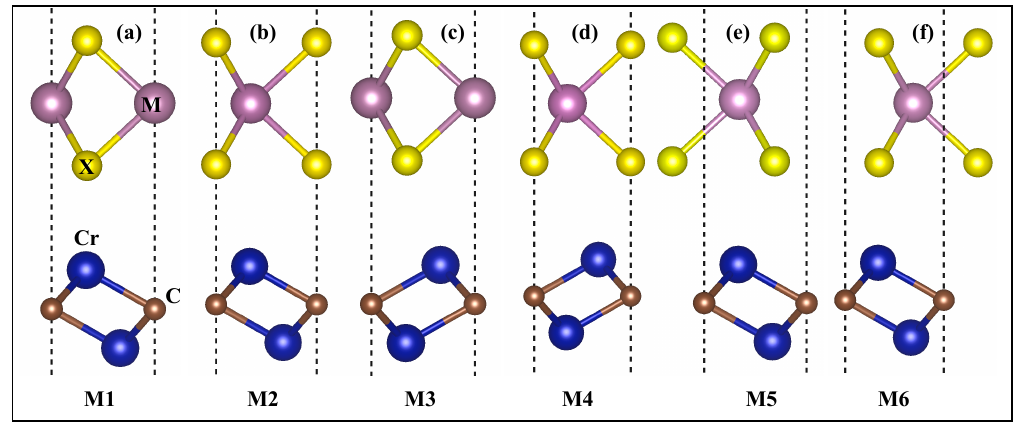}
    \caption{Side views of the six interface-stacking configurations considered in determining the most stable interface structure.}
    \label{fig:S1}
\end{figure}

Table~\ref{tab:stacking} lists the total energy differences $\Delta E = E_{\mathrm{M3}} - E_{\mathrm{M}i}$ ($i = 1$--$6$) of the six stacking configurations shown in Fig.~\ref{fig:S1}. For MoS$_2$ and MoSe$_2$, $\Delta E$ is negative for all configurations, indicating that M3 stacking is the most stable, whereas for WS$_2$ and WSe$_2$, the positive $\Delta E$ of M1 stacking shows that it is the most favorable.

\begin{table*}[h]
\caption{Calculated total energy differences $\Delta E = E_{\mathrm{M3}} - E_{\mathrm{M}i}$ (in meV) of the six stacking configurations for each Cr$_2$C/TMDC interface. A positive value indicates that the corresponding stacking is lower in energy than M3.}
\label{tab:stacking}
\begin{ruledtabular}
\begin{tabular}{lrrrr}
Stacking & MoS$_2$ & WS$_2$ & MoSe$_2$ & WSe$_2$ \\
\hline
M1 &  $-4.57$   &  $49.45$  &  $-3.54$  &  $16.66$  \\
M2 & $-122.80$  & $-105.83$ & $-63.85$  & $-59.91$  \\
M3 &   $0.00$   &   $0.00$  &   $0.00$  &   $0.00$  \\
M4 & $-58.42$   & $-68.04$  & $-29.07$  & $-11.02$  \\
M5 &  $-7.29$   &  $45.10$  &  $-5.20$  &  $15.15$  \\
M6 & $-38.91$   & $-78.13$  & $-47.60$  & $-58.87$  \\
\end{tabular}
\end{ruledtabular}
\end{table*}

\begin{figure}
    \centering
\includegraphics[width=16.0cm,clip]{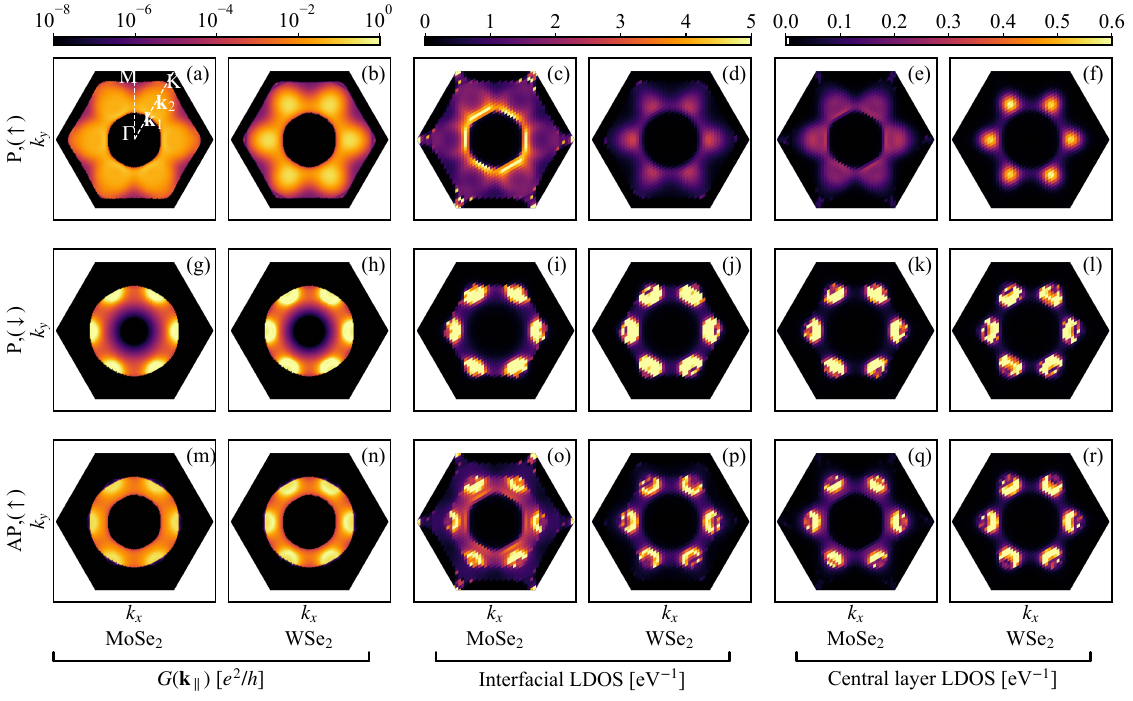}
\caption{$\mathbf{k}_{\parallel}$-resolved conductance at $E=E_\mathrm{F}$ for trilayer MoSe$_2$ [(a),(g),(m)] and WSe$_2$ [(b),(h),(n)]. For comparison the corresponding $\mathbf{k}_{\parallel}$-resolved LDOSs of the interfacial $Y$--$M$--$Y$ layers (sum of both left and right interfaces) are shown in columns [(c),(i),(o)] and [(d),(j),(p)], while those of the central $Y$--$M$--$Y$ layer in columns [(e),(k),(q)] and [(f),(l),(r)] for MoSe$_2$ and WSe$_2$, respectively. The top row [(a)–(f)] corresponds to the majority-spin channel and the middle row [(g)–(l)] to the minority-spin channel in the parallel (P) magnetization configuration. The bottom row [(m)–(r)] shows the majority-spin channel in the antiparallel (AP) magnetization configuration.}
\label{fig:kp3Lsm}
\end{figure}

\begin{figure}
    \centering
    \includegraphics[width=1\linewidth]{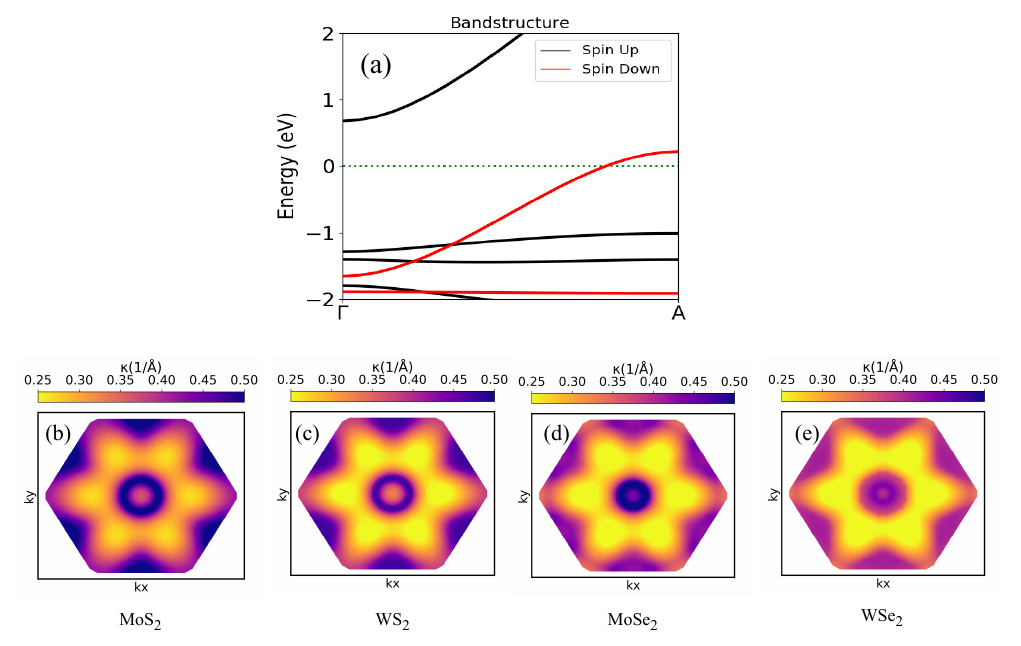}
    \caption{(a) Bulk band structure of the Cr$_2$C electrode along the transport direction ($\Gamma$--$A$). (b)--(e) Decay rates $\kappa$ of the evanescent states at the Fermi energy across the 2D Brillouin zone for (b) MoS$_2$, (c) WS$_2$, (d) MoSe$_2$, and (e) WSe$_2$. The values in $\kappa$ are divided by the thickness of the corresponding barrier.}
    \label{fig:S2}
\end{figure}

\begin{figure}
    \centering
\includegraphics[width=16.0cm,clip]{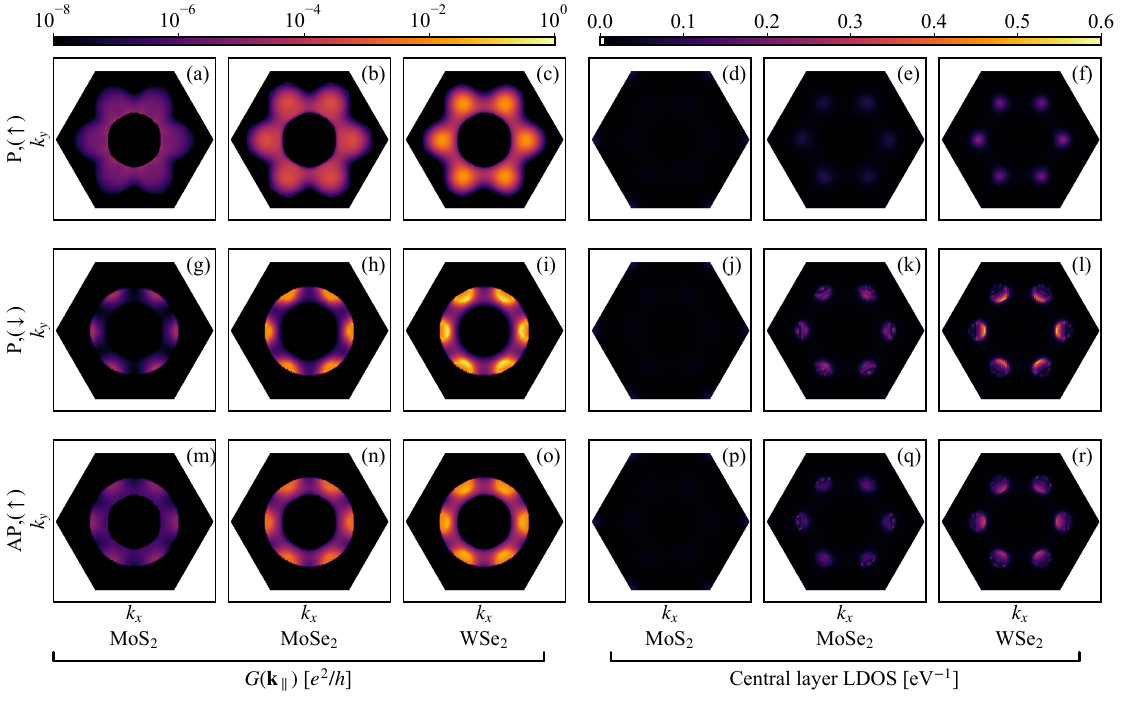}
\caption{$\mathbf{k}_{\parallel}$-resolved conductance $G(\mathbf{k}_{\parallel})$ at $E=E_\mathrm{F}$ for fivelayer MoS$_2$ [(a),(g),(m)], MoSe$_2$ [(b),(h),(n)] and WSe$_2$ [(c),(i),(o)]. For comparison the corresponding $\mathbf{k}_{\parallel}$-resolved LDOSs of the  central $Y$--$M$--$Y$ layer are shown in columns [(d),(j),(p)], [(e),(k),(q)] and [(f),(l),(r)] for MoS$_2$, MoSe$_2$ and WSe$_2$, respectively. The top row [(a)–(f)] corresponds to the majority-spin channel and the middle row [(g)–(l)] to the minority-spin channel in the parallel (P) magnetization configuration. The bottom row [(m)–(r)] shows the majority-spin channel in the antiparallel (AP) magnetization configuration.}
\label{fig:kpfivesm}
\end{figure}

\begin{figure}
\centering
\includegraphics[width=7.9cm, clip]{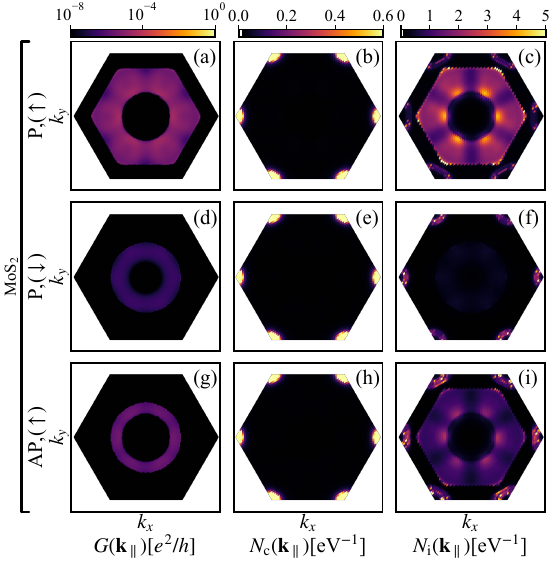}
\hfill
\includegraphics[width=7.9cm, clip]{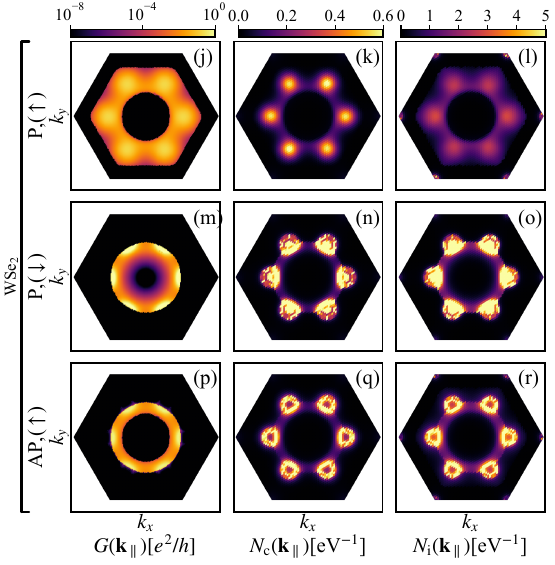}
\caption{$\mathbf{k}_{\parallel}$-resolved conductance $G(\mathbf{k}_{\parallel})$ at $E=E_\mathrm{F}$ and 4\% tensile strain for trilayer MoS$_2$ [(a),(d),(g)] and WSe$_2$ [(j),(m),(p)]. For comparison the corresponding $\mathbf{k}_{\parallel}$-resolved LDOSs of the central $Y$--$M$--$Y$ layer $N_c(\mathbf{k}_{\parallel})$ are shown in columns [(b),(e),(h)] and [(k),(n),(q)] while those of the interfacial $Y$--$M$--$Y$ layers (sum of both left and right interfaces) $N_i(\mathbf{k}_{\parallel})$ in columns [(c),(f),(i)] and [(l),(o),(r)] for MoS$_2$ and WSe$_2$, respectively. The top row [(a--c),(j--l)] corresponds to the majority-spin channel and the middle row [(d--f),(m--o)] to the minority-spin channel in the parallel (P) magnetization configuration. The bottom row [(g--i),(p--r)] shows the majority-spin channel in the antiparallel (AP) magnetization configuration.}
\label{fig:kpstrainsm}
\end{figure}

\begin{table}[h]
\caption{Calculated spin-resolved conductances and TMR ratios of the trilayer barriers with and without biaxial tensile strain. Conductances are in units of $10^{-3}\,e^2/h$, and TMR ratios in \%.}
\label{tab:tabstrain}
\begin{ruledtabular}
\begin{tabular}{lccccc c}
Barrier & Strain (\%) & \multicolumn{4}{c}{Conductance ($10^{-3}\,e^2/h$)} & TMR (\%) \\
\cline{3-6}
 & & $G_{\mathrm{P},\uparrow}$ & $G_{\mathrm{P},\downarrow}$ & $G_{\mathrm{AP},\uparrow}$ & $G_{\mathrm{AP},\downarrow}$ & \\
\hline
MoS$_2$ & 0 & 3.7 & 1.5 & 0.93 & 0.93 & 176 \\
 & 2 & 0.12 & 0.0044 & 0.017 & 0.017 & 266 \\
 & 3 & 0.082 & 0.0012 & 0.0075 & 0.0084 & 423 \\
 & 4 & 0.0032 & 0.000045 & 0.00027 & 0.00025 & 540 \\
\hline
WS$_2$ & 0 & 4.5 & 55 & 15 & 15 & 98 \\
 & 2 & 0.64 & 0.43 & 0.32 & 0.32 & 62 \\
 & 3 & 0.38 & 0.016 & 0.053 & 0.053 & 272 \\
 & 4 & 0.12 & 0.0024 & 0.010 & 0.010 & 496 \\
\hline
MoSe$_2$ & 0 & 12 & 39 & 12 & 12 & 110 \\
 & 2 & 5.5 & 17 & 5.0 & 5.0 & 120 \\
 & 3 & 4.8 & 6.7 & 2.4 & 2.4 & 143 \\
 & 4 & 3.4 & 1.9 & 1.0 & 1.0 & 162 \\
\hline
WSe$_2$ & 0 & 16 & 62 & 29 & 29 & 36 \\
 & 2 & 11 & 52 & 18 & 18 & 69 \\
 & 3 & 11 & 42 & 15 & 15 & 72 \\
 & 4 & 14 & 29 & 12 & 12 & 72 \\
\end{tabular}
\end{ruledtabular}
\end{table}